\documentclass[12pt]{article}
\usepackage{amsfonts}
\usepackage[dvips]{graphicx}
\usepackage{latexsym,amsmath,amssymb,graphics,stmaryrd}
\usepackage{array}
\usepackage{subfigure}
\usepackage{multirow}
\usepackage[dvips]{graphicx}
\usepackage[dvips]{color}
\usepackage{pstricks}
\usepackage{pst-node}
\usepackage{dsfont}
\usepackage{amsthm}
\usepackage{graphics}
\usepackage{subfigure}
\usepackage{fancyhdr}
\usepackage[dvips]{color}

\usepackage[textsize=tiny]{todonotes}

\makeatletter
\DeclareRobustCommand*{\bfseries}{%
\not@math@alphabet\bfseries\mathbf
\fontseries\bfdefault\selectfont
\boldmath
}
\makeatother

\usepackage[small,bf]{caption}

\usepackage{amsfonts}
\usepackage{bm}
\usepackage[dvips]{graphicx}
\usepackage{latexsym,amsmath,amssymb,graphics,stmaryrd}
\usepackage{cite}
\usepackage{array}
\usepackage{subfigure}
\usepackage{multirow}
\usepackage{epsfig}
\usepackage[dvips]{graphicx}
\usepackage[dvips]{color}
\usepackage{pstricks}
\usepackage{pst-node}
\usepackage{pst-plot}
\usepackage{amsthm}
\usepackage{graphics}
\usepackage{subfigure}
\usepackage{fancyhdr}

\usepackage{rotating}
\newlength{\arlength}
\newlength{\arheight}

\usepackage{feynmp}

\usepackage{feynmp}

\fancypagestyle{plain}{
\fancyhf{}
\newcommand{\fpage}{\iffloatpage{}{\thepage}}
\fancyfoot[C]{\fpage}

}

\newcommand{\col}{~,}
\newcommand{\pnt}{~.}
\newcommand{\AdS}{\text{AdS}}
\newcommand{\CFT}{\text{CFT}}

\newcommand{\twob}{{\text{II}\,\text{B}}}

\newcommand{\YM}{{\scriptscriptstyle\text{YM}}}

\newcommand{\deriv}[2][]{\frac{\de #1}{\de #2}}
\newcommand{\parderiv}[2][]{\frac{\partial #1}{\partial #2}}

\newcommand{\unitmatrix}{\mathds{1}}

\newcommand{\comm}[3][]{{}[{}#2{}\,{\overset{#1}{,}}\,{}#3{}]{}}

\newcommand{\de}{\operatorname{d}\!}

\newcommand{\e}{\operatorname{e}}

\newlength{\neglength}
\newlength{\diameter}

\newcommand{\svertex}[2]{%
\fmfiequ{#1}{point length(#2)/2 of (#2)}
}
\newcommand{\dvertex}[4][0.33]{%
\fmfiequ{#2}{point (0.5-(#1)/2)*length(#4) of (#4)}
\fmfiequ{#3}{point (0.5+(#1)/2)*length(#4) of (#4)}
}

\newcommand{\chionetwo}[1][black]{%
\fmftop{v1}
\fmfbottom{v4}
\fmfforce{(0.125w,h)}{v1}
\fmfforce{(0.125w,0)}{v4}
\fmffixed{(0.25w,0)}{v1,v2}
\fmffixed{(0.25w,0)}{v2,v3}
\fmffixed{(0.25w,0)}{v4,v5}
\fmffixed{(0.25w,0)}{v5,v6}
\fmffixed{(0,whatever)}{vc1,vc3}
\fmffixed{(0,whatever)}{vc2,vc4}
\fmf{plain,tension=0.5,right=0.25}{v1,vc1}
\fmf{plain,tension=0.5,left=0.25}{v2,vc1}
\fmf{phantom,tension=0.5,right=0.25}{v2,vc2}
\fmf{plain,tension=0.5,left=0.25}{v3,vc2}
\fmf{plain,tension=0.5,left=0.25}{v4,vc3}
\fmf{phantom,tension=0.5,right=0.25}{v5,vc3}
\fmf{plain,tension=0.5,left=0.25}{v5,vc4}
\fmf{plain,tension=0.5,right=0.25}{v6,vc4}
\fmf{plain,tension=1.25,left=0}{vc1,vc3}
\fmf{plain,tension=1.25,left=0}{vc2,vc4}
\fmffreeze
\fmf{plain,tension=1,left=0}{vc2,vc3}
\fmf{plain,tension=0.5,right=0,width=1mm}{v4,v6}
\fmffreeze
\fmfposition
\fmfipath{p[]}
\fmfipair{vd[],vm[],vu[]}
\fmfiset{p1}{vpath(__v1,__vc1)}
\fmfiset{p2}{vpath(__v2,__vc1)}
\fmfiset{p6}{vpath(__v3,__vc2)}
\fmfiset{p4}{vpath(__v4,__vc3)}
\fmfiset{p8}{vpath(__v5,__vc4)}
\fmfiset{p9}{vpath(__v6,__vc4)}
\fmfiset{p3}{vpath(__vc1,__vc3)}
\fmfiset{p7}{vpath(__vc2,__vc4)}
\fmfiset{p5}{vpath(__vc2,__vc3)}
\svertex{vm1}{p1}
\svertex{vm2}{p2}
\svertex{vm3}{p3}
\svertex{vm4}{p4}
\svertex{vm5}{p5}
\svertex{vm6}{p6}
\svertex{vm7}{p7}
\svertex{vm8}{p8}
\svertex{vm9}{p9}
}

\newcommand{\chionetwoone}[1][black]{%
\fmftop{v1}
\fmfbottom{v4}
\fmfforce{(0.125w,h)}{v1}
\fmfforce{(0.125w,0)}{v4}
\fmffixed{(0.25w,0)}{v1,v2}
\fmffixed{(0.25w,0)}{v2,v3}
\fmffixed{(0.25w,0)}{v4,v5}
\fmffixed{(0.25w,0)}{v5,v6}
\fmffixed{(0,whatever)}{vc1,vc3}
\fmffixed{(0,whatever)}{vb2,vb4}
\fmffixed{(0,whatever)}{vc1,vb1}
\fmffixed{(0,whatever)}{vc1,vb3}
\fmffixed{(whatever,0)}{vb1,vb2}
\fmffixed{(whatever,0)}{vb3,vb4}
\fmf{plain,tension=0.5,right=0.25}{v1,vc1}
\fmf{plain,tension=0.5,left=0.25}{v2,vc1}
\fmf{phantom,tension=0.5,right=0.25}{v2,vb2}
\fmf{plain,tension=0.5,left=0.25}{v3,vb2}
\fmf{plain,tension=0.5,left=0.25}{v4,vc3}
\fmf{plain,tension=0.5,right=0.25}{v5,vc3}
\fmf{phantom,tension=0.5,left=0.25}{v5,vb4}
\fmf{plain,tension=0.5,right=0.25}{v6,vb4}
\fmf{plain,tension=2,left=0}{vc1,vb1}
\fmf{plain,tension=2,left=0}{vb1,vb3}
\fmf{plain,tension=2,left=0}{vb3,vc3}
\fmf{plain,tension=2,left=0}{vb2,vb4}
\fmffreeze
\fmf{plain,tension=2,left=0}{vb1,vb2}
\fmf{plain,tension=2,left=0}{vb3,vb4}
\fmf{plain,tension=0.5,right=0,width=1mm}{v4,v6}
\fmffreeze
\fmfposition
\fmfipath{p[]}
\fmfipair{vd[],vm[],vu[]}
\fmfiset{p1}{vpath(__v1,__vc1)}
\fmfiset{p2}{vpath(__v2,__vc1)}
\fmfiset{p3}{vpath(__vc1,__vb1)}
\fmfiset{p4}{vpath(__vb1,__vb3)}
\fmfiset{p5}{vpath(__vb1,__vb2)}
\fmfiset{p6}{vpath(__v3,__vb2)}
\fmfiset{p7}{vpath(__vb2,__vb4)}
\fmfiset{p8}{vpath(__vb3,__vb4)}
\fmfiset{p9}{vpath(__v6,__vb4)}
\fmfiset{p10}{vpath(__vb3,__vc3)}
\fmfiset{p11}{vpath(__v4,__vc3)}
\fmfiset{p12}{vpath(__v5,__vc3)}
\svertex{vm1}{p1}
\svertex{vm2}{p2}
\svertex{vm3}{p3}
\svertex{vm4}{p4}
\svertex{vm5}{p5}
\svertex{vm6}{p6}
\svertex{vm7}{p7}
\svertex{vm8}{p8}
\svertex{vm9}{p9}
\svertex{vm10}{p10}
\svertex{vm11}{p11}
\svertex{vm12}{p12}
}

\newcommand{\chionetwothree}[1][black]{%
\fmftop{v1}
\fmfbottom{v5}
\fmfforce{(0.125w,h)}{v1}
\fmfforce{(0.125w,0)}{v5}
\fmffixed{(0.25w,0)}{v1,v2}
\fmffixed{(0.25w,0)}{v2,v3}
\fmffixed{(0.25w,0)}{v3,v4}
\fmffixed{(0.25w,0)}{v5,v6}
\fmffixed{(0.25w,0)}{v6,v7}
\fmffixed{(0.25w,0)}{v7,v8}
\fmffixed{(0,whatever)}{vc1,vc4}
\fmffixed{(0,whatever)}{vc2,vc5}
\fmffixed{(0,whatever)}{vc3,vc6}

\fmf{plain,tension=0.5,right=0.25}{v1,vc1}
\fmf{plain,tension=0.5,left=0.25}{v2,vc1}
\fmf{phantom,tension=0.5,right=0.25}{v2,vc2}
\fmf{plain,tension=0.5,left=0.25}{v3,vc2}
\fmf{phantom,tension=0.5,right=0.25}{v3,vc3}
\fmf{plain,tension=0.5,left=0.25}{v4,vc3}
\fmf{plain,tension=0.5,left=0.25}{v5,vc4}
\fmf{phantom,tension=0.5,right=0.25}{v6,vc4}
\fmf{plain,tension=0.5,left=0.25}{v6,vc5}
\fmf{phantom,tension=0.5,right=0.25}{v7,vc5}
\fmf{plain,tension=0.5,left=0.25}{v7,vc6}
\fmf{plain,tension=0.5,right=0.25}{v8,vc6}
\fmf{plain,tension=1.25,left=0}{vc1,vc4}
\fmf{plain,tension=1.25,left=0}{vc2,vc5}
\fmf{plain,tension=1.25,left=0}{vc3,vc6}
\fmffreeze
\fmf{plain,tension=1,left=0}{vc4,vc2}
\fmf{plain,tension=1,left=0}{vc5,vc3}
\fmf{plain,tension=0.5,right=0,width=1mm}{v5,v8}
\fmffreeze
\fmfposition
}

\newcommand{\chitwoonethree}[1][black]{%
\fmftop{v1}
\fmfbottom{v5}
\fmfforce{(0.125w,h)}{v1}
\fmfforce{(0.125w,0)}{v5}
\fmffixed{(0.25w,0)}{v1,v2}
\fmffixed{(0.25w,0)}{v2,v3}
\fmffixed{(0.25w,0)}{v3,v4}
\fmffixed{(0.25w,0)}{v5,v6}
\fmffixed{(0.25w,0)}{v6,v7}
\fmffixed{(0.25w,0)}{v7,v8}
\fmffixed{(whatever,0.5h)}{v5,vc1}
\fmffixed{(0,whatever)}{vc1,vc4}
\fmffixed{(0,whatever)}{vc2,vc5}
\fmffixed{(0,whatever)}{vc3,vc6}
\fmffixed{(whatever,0)}{vc1,vc3}
\fmffixed{(whatever,0)}{vc3,vc5}

\fmf{plain,tension=0.5,right=0.125}{v1,vc1}
\fmf{phantom,tension=0.5,left=0.25}{v2,vc1}
\fmf{plain,tension=0.5,right=0.25}{v2,vc2}
\fmf{plain,tension=0.5,left=0.25}{v3,vc2}
\fmf{phantom,tension=0.5,right=0.25}{v3,vc3}
\fmf{plain,tension=0.5,left=0.125}{v4,vc3}
\fmf{plain,tension=0.5,left=0.25}{v5,vc4}
\fmf{plain,tension=0.5,right=0.25}{v6,vc4}
\fmf{phantom,tension=0.5,left=0.25}{v6,vc5}
\fmf{phantom,tension=0.5,right=0.25}{v7,vc5}
\fmf{plain,tension=0.5,left=0.25}{v7,vc6}
\fmf{plain,tension=0.5,right=0.25}{v8,vc6}
\fmf{plain,tension=1.25,left=0}{vc1,vc4}
\fmf{plain,tension=1.25,left=0}{vc2,vc5}
\fmf{plain,tension=1.25,left=0}{vc3,vc6}
\fmffreeze
\fmf{plain,tension=1,left=0}{vc1,vc5}
\fmf{plain,tension=1,left=0}{vc5,vc3}
\fmf{plain,tension=0.5,right=0,width=1mm}{v5,v8}
\fmffreeze
\fmfposition
}

\newcommand{\chionethreetwo}[1][black]{%
\fmftop{v1}
\fmfbottom{v5}
\fmfforce{(0.125w,h)}{v1}
\fmfforce{(0.125w,0)}{v5}
\fmffixed{(0.25w,0)}{v1,v2}
\fmffixed{(0.25w,0)}{v2,v3}
\fmffixed{(0.25w,0)}{v3,v4}
\fmffixed{(0.25w,0)}{v5,v6}
\fmffixed{(0.25w,0)}{v6,v7}
\fmffixed{(0.25w,0)}{v7,v8}
\fmffixed{(whatever,0.5h)}{v5,vc2}
\fmffixed{(0,whatever)}{vc1,vc4}
\fmffixed{(0,whatever)}{vc2,vc5}
\fmffixed{(0,whatever)}{vc3,vc6}
\fmffixed{(whatever,0)}{vc2,vc4}
\fmffixed{(whatever,0)}{vc4,vc6}
\fmf{plain,tension=0.5,right=0.25}{v1,vc1}
\fmf{plain,tension=0.5,left=0.25}{v2,vc1}
\fmf{phantom,tension=0.5,right=0.25}{v2,vc2}
\fmf{phantom,tension=0.5,left=0.25}{v3,vc2}
\fmf{plain,tension=0.5,right=0.25}{v3,vc3}
\fmf{plain,tension=0.5,left=0.25}{v4,vc3}
\fmf{plain,tension=0.5,left=0.125}{v5,vc4}
\fmf{phantom,tension=0.5,right=0.25}{v6,vc4}
\fmf{plain,tension=0.5,left=0.25}{v6,vc5}
\fmf{plain,tension=0.5,right=0.25}{v7,vc5}
\fmf{phantom,tension=0.5,left=0.25}{v7,vc6}
\fmf{plain,tension=0.5,right=0.125}{v8,vc6}
\fmf{plain,tension=1.25,left=0}{vc1,vc4}
\fmf{plain,tension=1.25,left=0}{vc2,vc5}
\fmf{plain,tension=1.25,left=0}{vc3,vc6}
\fmffreeze
\fmf{plain,tension=1,left=0}{vc2,vc4}
\fmf{plain,tension=1,left=0}{vc2,vc6}
\fmf{plain,tension=0.5,right=0,width=1mm}{v5,v8}
\fmffreeze
\fmfposition
}

\newcommand{\chionetwothreetwo}[1][black]{%
\fmftop{v1}
\fmfbottom{v5}
\fmfforce{(0.125w,h)}{v1}
\fmfforce{(0.125w,0)}{v5}
\fmffixed{(0.25w,0)}{v1,v2}
\fmffixed{(0.25w,0)}{v2,v3}
\fmffixed{(0.25w,0)}{v3,v4}
\fmffixed{(0.25w,0)}{v5,v6}
\fmffixed{(0.25w,0)}{v6,v7}
\fmffixed{(0.25w,0)}{v7,v8}
\fmffixed{(0,whatever)}{va1,va2}
\fmffixed{(whatever,0)}{va1,vc1}
\fmffixed{(whatever,0)}{va2,vb3}
\fmffixed{(0,whatever)}{vc1,vc3}
\fmffixed{(0,whatever)}{vb2,vb4}
\fmffixed{(0,whatever)}{vc1,vb1}
\fmffixed{(0,whatever)}{vc1,vb3}
\fmffixed{(whatever,0)}{vb1,vb2}
\fmffixed{(whatever,0)}{vb3,vb4}
\fmf{phantom,tension=0.5,right=0.25}{v2,vc1}
\fmf{plain,tension=0.5,left=0.25}{v3,vc1}
\fmf{phantom,tension=0.5,right=0.25}{v3,vb2}
\fmf{plain,tension=0.5,left=0.25}{v4,vb2}
\fmf{plain,tension=0.5,left=0.25}{v6,vc3}
\fmf{plain,tension=0.5,right=0.25}{v7,vc3}
\fmf{phantom,tension=0.5,left=0.25}{v7,vb4}
\fmf{plain,tension=0.5,right=0.25}{v8,vb4}
\fmf{plain,tension=2,left=0}{vc1,vb1}
\fmf{plain,tension=2,left=0}{vb1,vb3}
\fmf{plain,tension=2,left=0}{vb3,vc3}
\fmf{plain,tension=2,left=0}{vb2,vb4}
\fmffreeze
\fmf{plain,tension=0.5,right=0.25}{v1,va1}
\fmf{plain,tension=0.5,left=0.25}{v2,va1}
\fmf{plain,tension=1}{va1,va2}
\fmf{plain,tension=0}{va2,vc1}
\fmf{plain,tension=0.5,left=0.25}{v5,va2}
\fmf{phantom,tension=0.5,right=0.25}{v6,va2}
\fmf{plain,tension=1,left=0}{vb1,vb2}
\fmf{plain,tension=1,left=0}{vb3,vb4}
\fmf{plain,tension=0.5,right=0,width=1mm}{v5,v8}
}

\newcommand{\chitwothreetwoone}[1][black]{%
\fmftop{v1}
\fmfbottom{v5}
\fmfforce{(0.125w,h)}{v1}
\fmfforce{(0.125w,0)}{v5}
\fmffixed{(0.25w,0)}{v1,v2}
\fmffixed{(0.25w,0)}{v2,v3}
\fmffixed{(0.25w,0)}{v3,v4}
\fmffixed{(0.25w,0)}{v5,v6}
\fmffixed{(0.25w,0)}{v6,v7}
\fmffixed{(0.25w,0)}{v7,v8}
\fmffixed{(0,whatever)}{va1,va2}
\fmffixed{(whatever,0)}{va2,vc3}
\fmffixed{(whatever,0)}{va1,vb1}
\fmffixed{(0,whatever)}{vc1,vc3}
\fmffixed{(0,whatever)}{vb2,vb4}
\fmffixed{(0,whatever)}{vc1,vb1}
\fmffixed{(0,whatever)}{vc1,vb3}
\fmffixed{(whatever,0)}{vb1,vb2}
\fmffixed{(whatever,0)}{vb3,vb4}
\fmf{plain,tension=0.5,right=0.25}{v2,vc1}
\fmf{plain,tension=0.5,left=0.25}{v3,vc1}
\fmf{phantom,tension=0.5,right=0.25}{v3,vb2}
\fmf{plain,tension=0.5,left=0.25}{v4,vb2}
\fmf{phantom,tension=0.5,left=0.25}{v6,vc3}
\fmf{plain,tension=0.5,right=0.25}{v7,vc3}
\fmf{phantom,tension=0.5,left=0.25}{v7,vb4}
\fmf{plain,tension=0.5,right=0.25}{v8,vb4}
\fmf{plain,tension=2,left=0}{vc1,vb1}
\fmf{plain,tension=2,left=0}{vb1,vb3}
\fmf{plain,tension=2,left=0}{vb3,vc3}
\fmf{plain,tension=2,left=0}{vb2,vb4}
\fmffreeze
\fmf{plain,tension=0.5,right=0.25}{v1,va1}
\fmf{phantom,tension=0.5,left=0.25}{v2,va1}
\fmf{plain,tension=1}{va1,va2}
\fmf{plain,tension=0}{va1,vc3}
\fmf{plain,tension=0.5,left=0.25}{v5,va2}
\fmf{plain,tension=0.5,right=0.25}{v6,va2}
\fmf{plain,tension=1,left=0}{vb1,vb2}
\fmf{plain,tension=1,left=0}{vb3,vb4}
\fmf{plain,tension=0.5,right=0,width=1mm}{v5,v8}
}

\newcommand{\chitwoonetwothree}[1][black]{%
\fmftop{v1}
\fmfbottom{v5}
\fmfforce{(0.125w,h)}{v1}
\fmfforce{(0.125w,0)}{v5}
\fmffixed{(0.25w,0)}{v1,v2}
\fmffixed{(0.25w,0)}{v2,v3}
\fmffixed{(0.25w,0)}{v3,v4}
\fmffixed{(0.25w,0)}{v5,v6}
\fmffixed{(0.25w,0)}{v6,v7}
\fmffixed{(0.25w,0)}{v7,v8}
\fmffixed{(0,whatever)}{va1,va2}
\fmffixed{(whatever,0)}{va2,vc3}
\fmffixed{(whatever,0)}{va1,vb1}
\fmffixed{(0,whatever)}{vc1,vc3}
\fmffixed{(0,whatever)}{vb2,vb4}
\fmffixed{(0,whatever)}{vc1,vb1}
\fmffixed{(0,whatever)}{vc1,vb3}
\fmffixed{(whatever,0)}{vb1,vb2}
\fmffixed{(whatever,0)}{vb3,vb4}
\fmf{plain,tension=0.5,right=0.25}{v2,vc1}
\fmf{plain,tension=0.5,left=0.25}{v3,vc1}
\fmf{plain,tension=0.5,right=0.25}{v1,vb2}
\fmf{phantom,tension=0.5,left=0.25}{v2,vb2}
\fmf{plain,tension=0.5,left=0.25}{v6,vc3}
\fmf{phantom,tension=0.5,right=0.25}{v7,vc3}
\fmf{plain,tension=0.5,left=0.25}{v5,vb4}
\fmf{phantom,tension=0.5,right=0.25}{v6,vb4}
\fmf{plain,tension=2,left=0}{vc1,vb1}
\fmf{plain,tension=2,left=0}{vb1,vb3}
\fmf{plain,tension=2,left=0}{vb3,vc3}
\fmf{plain,tension=2,left=0}{vb2,vb4}
\fmffreeze
\fmf{phantom,tension=0.5,right=0.25}{v3,va1}
\fmf{plain,tension=0.5,left=0.25}{v4,va1}
\fmf{plain,tension=1}{va1,va2}
\fmf{plain,tension=0}{va1,vc3}
\fmf{plain,tension=0.5,left=0.25}{v7,va2}
\fmf{plain,tension=0.5,right=0.25}{v8,va2}
\fmf{plain,tension=1,left=0}{vb1,vb2}
\fmf{plain,tension=1,left=0}{vb3,vb4}
\fmf{plain,tension=0.5,right=0,width=1mm}{v5,v8}
}

\newcommand{\chithreeonetwoone}[1][black]{%
\fmftop{v1}
\fmfbottom{v5}
\fmfforce{(0.125w,h)}{v1}
\fmfforce{(0.125w,0)}{v5}
\fmffixed{(0.25w,0)}{v1,v2}
\fmffixed{(0.25w,0)}{v2,v3}
\fmffixed{(0.25w,0)}{v3,v4}
\fmffixed{(0.25w,0)}{v5,v6}
\fmffixed{(0.25w,0)}{v6,v7}
\fmffixed{(0.25w,0)}{v7,v8}
\fmffixed{(0,whatever)}{va1,va2}
\fmffixed{(whatever,0)}{va1,vc1}
\fmffixed{(whatever,0)}{va2,vb1}
\fmffixed{(0,whatever)}{vc1,vc3}
\fmffixed{(0,whatever)}{vb2,vb4}
\fmffixed{(0,whatever)}{vc1,vb1}
\fmffixed{(0,whatever)}{vc1,vb3}
\fmffixed{(whatever,0)}{vb1,vb2}
\fmffixed{(whatever,0)}{vb3,vb4}
\fmf{plain,tension=0.5,right=0.25}{v1,vc1}
\fmf{plain,tension=0.5,left=0.25}{v2,vc1}
\fmf{phantom,tension=0.5,right=0.25}{v2,vb2}
\fmf{phantom,tension=0.5,left=0.25}{v3,vb2}
\fmf{plain,tension=0.5,left=0.25}{v5,vc3}
\fmf{plain,tension=0.5,right=0.25}{v6,vc3}
\fmf{phantom,tension=0.5,left=0.125}{v6,vb4}
\fmf{plain,tension=0.5,right=0.125}{v7,vb4}
\fmf{plain,tension=2,left=0}{vc1,vb1}
\fmf{plain,tension=2,left=0}{vb1,vb3}
\fmf{plain,tension=2,left=0}{vb3,vc3}
\fmf{plain,tension=2,left=0}{vb2,vb4}
\fmffreeze
\fmf{plain,tension=0.5,right=0.25}{v3,va1}
\fmf{plain,tension=0.5,left=0.25}{v4,va1}
\fmf{plain,tension=1}{va1,va2}
\fmf{plain,tension=0}{va2,vb2}
\fmf{phantom,tension=0.5,left=0.125}{v7,va2}
\fmf{plain,tension=0.5,right=0.125}{v8,va2}
\fmf{plain,tension=1,left=0}{vb1,vb2}
\fmf{plain,tension=1,left=0}{vb3,vb4}
\fmf{plain,tension=0.5,right=0,width=1mm}{v5,v8}
}

\newcommand{\chionethreetwothree}[1][black]{%
\fmftop{v1}
\fmfbottom{v5}
\fmfforce{(0.125w,h)}{v1}
\fmfforce{(0.125w,0)}{v5}
\fmffixed{(0.25w,0)}{v1,v2}
\fmffixed{(0.25w,0)}{v2,v3}
\fmffixed{(0.25w,0)}{v3,v4}
\fmffixed{(0.25w,0)}{v5,v6}
\fmffixed{(0.25w,0)}{v6,v7}
\fmffixed{(0.25w,0)}{v7,v8}
\fmffixed{(0,whatever)}{va1,va2}
\fmffixed{(whatever,0)}{va1,vc1}
\fmffixed{(whatever,0)}{va2,vb1}
\fmffixed{(0,whatever)}{vc1,vc3}
\fmffixed{(0,whatever)}{vb2,vb4}
\fmffixed{(0,whatever)}{vc1,vb1}
\fmffixed{(0,whatever)}{vc1,vb3}
\fmffixed{(whatever,0)}{vb1,vb2}
\fmffixed{(whatever,0)}{vb3,vb4}
\fmf{plain,tension=0.5,right=0.25}{v3,vc1}
\fmf{plain,tension=0.5,left=0.25}{v4,vc1}
\fmf{phantom,tension=0.5,right=0.25}{v2,vb2}
\fmf{phantom,tension=0.5,left=0.25}{v3,vb2}
\fmf{plain,tension=0.5,left=0.25}{v7,vc3}
\fmf{plain,tension=0.5,right=0.25}{v8,vc3}
\fmf{plain,tension=0.5,left=0.125}{v6,vb4}
\fmf{phantom,tension=0.5,right=0.125}{v7,vb4}
\fmf{plain,tension=2,left=0}{vc1,vb1}
\fmf{plain,tension=2,left=0}{vb1,vb3}
\fmf{plain,tension=2,left=0}{vb3,vc3}
\fmf{plain,tension=2,left=0}{vb2,vb4}
\fmffreeze
\fmf{plain,tension=0.5,right=0.25}{v1,va1}
\fmf{plain,tension=0.5,left=0.25}{v2,va1}
\fmf{plain,tension=1}{va1,va2}
\fmf{plain,tension=0}{va2,vb2}
\fmf{plain,tension=0.5,left=0.125}{v5,va2}
\fmf{phantom,tension=0.5,right=0.125}{v6,va2}
\fmf{plain,tension=1,left=0}{vb1,vb2}
\fmf{plain,tension=1,left=0}{vb3,vb4}
\fmf{plain,tension=0.5,right=0,width=1mm}{v5,v8}
}

\newcommand{\chionetwoonethree}[1][black]{%
\fmftop{v1}
\fmfbottom{v5}
\fmfforce{(0.125w,h)}{v1}
\fmfforce{(0.125w,0)}{v5}
\fmffixed{(0.25w,0)}{v1,v2}
\fmffixed{(0.25w,0)}{v2,v3}
\fmffixed{(0.25w,0)}{v3,v4}
\fmffixed{(0.25w,0)}{v5,v6}
\fmffixed{(0.25w,0)}{v6,v7}
\fmffixed{(0.25w,0)}{v7,v8}
\fmffixed{(0,whatever)}{va1,va2}
\fmffixed{(whatever,0)}{va1,vb3}
\fmffixed{(whatever,0)}{va2,vc3}
\fmffixed{(0,whatever)}{vc1,vc3}
\fmffixed{(0,whatever)}{vb2,vb4}
\fmffixed{(0,whatever)}{vc1,vb1}
\fmffixed{(0,whatever)}{vc1,vb3}
\fmffixed{(whatever,0)}{vb1,vb2}
\fmffixed{(whatever,0)}{vb3,vb4}
\fmf{plain,tension=0.5,right=0.25}{v1,vc1}
\fmf{plain,tension=0.5,left=0.25}{v2,vc1}
\fmf{phantom,tension=0.5,right=0.125}{v2,vb2}
\fmf{plain,tension=0.5,left=0.125}{v3,vb2}
\fmf{plain,tension=0.5,left=0.25}{v5,vc3}
\fmf{plain,tension=0.5,right=0.25}{v6,vc3}
\fmf{phantom,tension=0.5,left=0.25}{v6,vb4}
\fmf{phantom,tension=0.5,right=0.25}{v7,vb4}
\fmf{plain,tension=2,left=0}{vc1,vb1}
\fmf{plain,tension=2,left=0}{vb1,vb3}
\fmf{plain,tension=2,left=0}{vb3,vc3}
\fmf{plain,tension=2,left=0}{vb2,vb4}
\fmffreeze
\fmf{phantom,tension=0.5,right=0.125}{v3,va1}
\fmf{plain,tension=0.5,left=0.125}{v4,va1}
\fmf{plain,tension=1}{va1,va2}
\fmf{plain,tension=0}{va1,vb4}
\fmf{plain,tension=0.5,left=0.25}{v7,va2}
\fmf{plain,tension=0.5,right=0.25}{v8,va2}
\fmf{plain,tension=1,left=0}{vb1,vb2}
\fmf{plain,tension=1,left=0}{vb3,vb4}
\fmf{plain,tension=0.5,right=0,width=1mm}{v5,v8}
}

\newcommand{\chionetwoonetwoone}[1][black]{%
\fmftop{v1}
\fmfbottom{v4}
\fmfforce{(0.125w,h)}{v1}
\fmfforce{(0.125w,0)}{v4}
\fmffixed{(0.25w,0)}{v1,v2}
\fmffixed{(0.25w,0)}{v2,v3}
\fmffixed{(0.25w,0)}{v4,v5}
\fmffixed{(0.25w,0)}{v5,v6}
\fmffixed{(0,whatever)}{vc1,vc3}
\fmffixed{(0,whatever)}{vb2,vb4}
\fmffixed{(0,whatever)}{vc2,vb2}
\fmffixed{(0,whatever)}{vc4,vb4}
\fmffixed{(0,whatever)}{vc1,vb1}
\fmffixed{(0,whatever)}{vc1,vb3}
\fmffixed{(whatever,0)}{vc1,vc2}
\fmffixed{(whatever,0)}{vb1,vb2}
\fmffixed{(whatever,0)}{vb3,vb4}
\fmffixed{(whatever,0)}{vc3,vc4}
\fmf{plain,tension=0.5,right=0.25}{v1,vc1}
\fmf{plain,tension=0.5,left=0.25}{v2,vc1}
\fmf{phantom,tension=0.5,right=0.25}{v2,vc2}
\fmf{plain,tension=0.5,left=0.25}{v3,vc2}
\fmf{plain,tension=0.5,left=0.25}{v4,vc3}
\fmf{phantom,tension=0.5,right=0.25}{v5,vc3}
\fmf{plain,tension=0.5,left=0.25}{v5,vc4}
\fmf{plain,tension=0.5,right=0.25}{v6,vc4}
\fmf{plain,tension=2,left=0}{vc1,vb1}
\fmf{plain,tension=2,left=0}{vb1,vb3}
\fmf{plain,tension=2,left=0}{vb3,vc3}
\fmf{plain,tension=2,left=0}{vc2,vb2}
\fmf{plain,tension=2,left=0}{vb2,vb4}
\fmf{plain,tension=2,left=0}{vb4,vc4}
\fmffreeze
\fmf{plain,tension=2,left=0}{vb1,vc2}
\fmf{plain,tension=2,left=0}{vb3,vb2}
\fmf{plain,tension=2,left=0}{vc3,vb4}
\fmf{plain,tension=0.5,right=0,width=1mm}{v4,v6}
}

\newcommand{\nvml}[3][1]{%
\fmfcmd{%
begingroup;
save a, vp, tvp, nvp, tv, nv, ip, ts, tt, is, it, n, m, scale, t, r, s, ttpr, tnpr, ep, mm;
path lcirc;
pair vp[][], tvp[][], tv[][], nvp[][], nv[][], ip[][], ts[], is[], tt[], it[], ep[], mid;
n := #2;
m:=3;
for i=1 upto n:
for j=1 upto m:
a[i][j] := arctime ((j-1)/(m-1)*arclength pm[i]) of pm[i];
vp[i][j] := point a[i][j] of pm[i];
tvp[i][j] := unitvector direction a[i][j] of pm[i];
nvp[i][j] := tvp[i][j] rotated -90;
endfor;
endfor;
if(vp[1][1]=vp[n][m]):
vp[n+1][1] := vp[1][1];
tvp[0][m] := tvp[n][m];
nvp[0][m] := nvp[n][m];
tvp[n+1][1] :=tvp[1][1];
nvp[n+1][1] :=nvp[1][1];
else:
vp[n+1][1] := vp[n][m];
tvp[0][m] := (0,0);
nvp[0][m] := (0,0);
tvp[n+1][1] :=tvp[n][m];
nvp[n+1][1] :=nvp[n][m];
fi;
s := 1;
for i=1 upto n:
for j=1 upto m:
if (j=1):
tv[i][1] := (tvp[i-1][m]+tvp[i][1]);
nv[i][1] := (nvp[i-1][m]+nvp[i][1]);
if (not(tv[i][1]=(0,0))):
tv[i][1] := unitvector tv[i][1];
fi;
if (not(nv[i][1]=(0,0))):
nv[i][1] := unitvector nv[i][1];
fi;
ttpr := tvp[i][1] dotprod tvp[i-1][m];
tnpr := tvp[i][1] dotprod nvp[i-1][m];
elseif (j=m):
tv[i][m] := (tvp[i][m]+tvp[i+1][1]);
nv[i][m] := (nvp[i][m]+nvp[i+1][1]);
if (not(tv[i][m]=(0,0))):
tv[i][m] := unitvector tv[i][m];
fi;
if (not(nv[i][m]=(0,0))):
nv[i][m] := unitvector nv[i][m];
fi;
ttpr := tvp[i][m] dotprod tvp[i+1][1];
tnpr := -tvp[i][m] dotprod nvp[i+1][1];
else:
nv[i][j] :=nvp[i][j];
tv[i][j] :=tvp[i][j];
fi;
scale := 25;
if ((j=1) or (j=m)):
 if ((tnpr<=0) and not((tv[i][j]=(0,0)) or (nv[i][j]=(0,0)))):
  ip[i][j] := vp[i][j] shifted(0.15*scale*nvp[i][j]);
  ts[s] := tvp[i][j];
  is[s] := ip[i][j];
  s:=s+1;
 else:
  if ((j=1) and (ttpr>0)):
  fi;
 fi;
else:
 ip[i][j] := vp[i][j] shifted(0.15*scale*nv[i][j]);
 ts[s] := tv[i][j];
 is[s] := ip[i][j];
 s:=s+1;
fi;
endfor;
endfor;
if(vp[1][1]=vp[n][m]):
ts[s] := ts[1];
is[s] := is[1];
else:
tv[n+1][1] := unitvector (tvp[n][m]+tvp[n+1][1]);
nv[n+1][1] := unitvector (nvp[n][m]+nvp[n+1][1]);
ip[n+1][1] := vp[n+1][1] shifted(0.15*scale*nv[n+1][1]);
ts[s] := tv[n+1][1];
is[s] := ip[n+1][1];
fi;
t=#1;
lcirc:=is[1];
for k=2 upto s:
lcirc := lcirc{ts[k-1]}..tension t..{ts[k]}is[k];
endfor;
mm := arctime (0.5* arclength lcirc) of lcirc;
if(vp[1][1]=vp[n][m]):
ep1 := point arctime (0* arclength lcirc) of lcirc of lcirc;
ep2 := point mm of lcirc;
mid := 1/2[ep1,ep2];
else:
ep1 := point mm of lcirc;
ep2 :=unitvector direction mm of lcirc rotated -90;
mid:= ep1 shifted(0.2*scale*ep2);
fi;
draw(lcirc) withpen pencircle scaled 0.25;
drawarrow(subpath(mm*0.8,mm*1.1) of lcirc) withpen pencircle scaled 0.25;
endgroup;
}
\fmfiv{label=#3,l.dist=0}{mid}
}

\DeclareMathOperator{\tr}{tr}

\DeclareMathOperator{\T}{T}

\DeclareMathOperator{\Top}{T}
\DeclareMathOperator{\Kop}{K}
\DeclareMathOperator{\Rop}{R}

\DeclareMathOperator{\D}{D}

\newlength{\eqoff}

\newlength{\unit}
\psset{xunit=\unit,yunit=\unit,runit=\unit}
\newlength{\linew}
\psset{linewidth=\linew}

\usepackage[a4paper,top=3cm,bottom=3cm,left=3cm,right=3cm]{geometry}

\numberwithin{equation}{section}

\usepackage{hyperref}

\newgray{ogray}{0.85}
\newgray{hatchgray}{1}
\newgray{sgray}{0.8}

\newcommand{\sfillstyle}{solid}

\newlength{\rad}
\newlength{\roff}
\newlength{\ri}
\psset{xunit=\unit,yunit=\unit,runit=\unit}
\newlength{\dlinewidth}
\newlength{\doublesep}
\psset{doublesep=\doublesep}
\psset{linewidth=\linew}
\newlength{\auxlen}
\newlength{\linearc}
\newlength{\xa}
\newlength{\ya}
\newlength{\xb}
\newlength{\yb}
\newlength{\xc}
\newlength{\yc}
\newlength{\xd}
\newlength{\yd}

\newlength{\ye}

\newlength{\yf}

\newcommand{\threevertex}[4][white]{%
\setlength{\xa}{0\unit}
\addtolength{\xa}{-0.5\dlinewidth}
\setlength{\xb}{0\unit}
\addtolength{\xb}{0.5\dlinewidth}
\setlength{\xc}{0\unit}
\addtolength{\xc}{1\unit}
\setlength{\ya}{0\unit}
\addtolength{\ya}{1\unit}
\setlength{\yb}{0\unit}
\addtolength{\yb}{0.5\dlinewidth}
\setlength{\yc}{0\unit}
\addtolength{\yc}{-0.5\dlinewidth}
\setlength{\yd}{0\unit}
\addtolength{\yd}{-1\unit}
\psset{doubleline=false}
\rput{0}(#2\unit,#3\unit){%
\pscustom[fillstyle=\sfillstyle,fillcolor=#1,linecolor=#1,linewidth=0pt]{%
\rotate{#4}
\psline[liftpen=1,linearc=\linearc](\xb,\ya)(\xb,\yb)(\xc,\yb)
\psline[liftpen=1,linearc=\linearc](\xc,\yc)(\xb,\yc)(\xb,\yd)
\psline[liftpen=1](\xa,\yd)(\xa,\ya)}
\pscustom{%
\rotate{#4}
\psline[liftpen=2,linearc=\linearc](\xb,\ya)(\xb,\yb)(\xc,\yb)
\psline[liftpen=2,linearc=\linearc](\xc,\yc)(\xb,\yc)(\xb,\yd)
\psline[liftpen=2](\xa,\yd)(\xa,\ya)
}
}
}

\newcommand{\recthreevertex}[4][white]{%
\setlength{\xa}{-0.7071\auxlen}
\addtolength{\xa}{-0.7071\dlinewidth}
\setlength{\xb}{0\unit}
\addtolength{\xb}{-0.7071\auxlen}
\setlength{\xc}{-0.7071\auxlen}
\addtolength{\xc}{0.7071\unit}
\setlength{\xd}{-0.7071\auxlen}
\addtolength{\xd}{1.2071\unit}

\setlength{\ya}{-0.7071\auxlen}
\addtolength{\ya}{-0.7071\dlinewidth}
\setlength{\yb}{0\unit}
\addtolength{\yb}{-0.7071\auxlen}
\setlength{\yc}{-0.7071\auxlen}
\addtolength{\yc}{-0.7071\dlinewidth}
\addtolength{\yc}{0.7071\unit}
\setlength{\yd}{0.7071\auxlen}
\addtolength{\yd}{0.7071\dlinewidth}
\addtolength{\yd}{-0.7071\unit}
\setlength{\ye}{0\unit}
\addtolength{\ye}{0.7071\auxlen}
\setlength{\yf}{0.7071\auxlen}
\addtolength{\yf}{0.7071\dlinewidth}
\psset{doubleline=false}
\rput{0}(#2\unit,#3\unit){%
\pscustom[fillstyle=\sfillstyle,fillcolor=#1,linecolor=#1,linewidth=0pt]{%
\rotate{#4}
\psline[liftpen=1,linearc=\linearc](\xa,\yb)(-0.7071\dlinewidth,0)(\xa,\ye)
\psline[liftpen=1,linearc=\linearc](\xb,\yf)(\xc,\yd)(\xd,\yd)
\psline[liftpen=1,linearc=\linearc](\xd,\yc)(\xc,\yc)(\xb,\ya)}
\pscustom{%
\rotate{#4}
\psline[liftpen=1,linearc=\linearc](\xa,\yb)(-0.7071\dlinewidth,0)(\xa,\ye)
\psline[liftpen=2,linearc=\linearc](\xb,\yf)(\xc,\yd)(\xd,\yd)
\psline[liftpen=2,linearc=\linearc](\xd,\yc)(\xc,\yc)(\xb,\ya)
}
}
}

\newcommand{\recthreevertexoneside}[4][white]{%
\setlength{\xa}{-0.7071\auxlen}
\addtolength{\xa}{-0.7071\dlinewidth}
\setlength{\xb}{0\unit}
\addtolength{\xb}{-0.7071\auxlen}
\setlength{\xc}{-0.7071\auxlen}
\addtolength{\xc}{-0.353505\unit}
\setlength{\xd}{-\auxlen}
\addtolength{\xd}{-0.5\dlinewidth}
\setlength{\ya}{-0.7071\auxlen}
\addtolength{\ya}{-0.7071\dlinewidth}
\setlength{\yb}{0\unit}
\addtolength{\yb}{-0.7071\auxlen}
\setlength{\yc}{-0.7071\auxlen}
\addtolength{\yc}{-0.7071\dlinewidth}
\addtolength{\yc}{0.7071\unit}
\setlength{\yd}{0.7071\auxlen}
\addtolength{\yd}{0.7071\dlinewidth}
\addtolength{\yd}{-0.7071\unit}
\setlength{\ye}{0\unit}
\addtolength{\ye}{0.7071\auxlen}
\setlength{\yf}{0.7071\auxlen}
\addtolength{\yf}{0.7071\dlinewidth}
\psset{doubleline=false}
\rput{0}(#2\unit,#3\unit){%
\pscustom[fillstyle=\sfillstyle,fillcolor=#1,linecolor=#1,linewidth=0pt]{%
\rotate{#4}
\psline[liftpen=1,linearc=\linearc](\xb,\yf)(0.7071\dlinewidth,0)(\xb,\ya)
\psline[liftpen=1,linearc=0.2\linearc](\xa,\yb)(\xc,\yc)(\xd,\yc)
\psline[liftpen=1,linearc=0.2\linearc](\xd,\yd)(\xc,\yd)(\xa,\ye)}
\pscustom{%
\rotate{#4}
\psline[liftpen=1,linearc=\linearc](\xb,\yf)(0.7071\dlinewidth,0)(\xb,\ya)
\psline[liftpen=2,linearc=0.2\linearc](\xa,\yb)(\xc,\yc)(\xd,\yc)
\psline[liftpen=2,linearc=0.2\linearc](\xd,\yd)(\xc,\yd)(\xa,\ye)
}
}
}
\newcommand{\fourvertex}[4][white]{%
\setlength{\xa}{0\unit}
\addtolength{\xa}{-1\unit}
\setlength{\xb}{0\unit}
\addtolength{\xb}{-0.5\dlinewidth}
\setlength{\xc}{0\unit}
\addtolength{\xc}{0.5\dlinewidth}
\setlength{\xd}{0\unit}
\addtolength{\xd}{1\unit}
\setlength{\ya}{0\unit}
\addtolength{\ya}{-1\unit}
\setlength{\yb}{0\unit}
\addtolength{\yb}{-0.5\dlinewidth}
\setlength{\yc}{0\unit}
\addtolength{\yc}{0.5\dlinewidth}
\setlength{\yd}{0\unit}
\addtolength{\yd}{1\unit}
\psset{doubleline=false}
\rput{0}(#2\unit,#3\unit){%
\pscustom[fillstyle=\sfillstyle,fillcolor=#1,linecolor=#1,linewidth=0pt]{%
\rotate{#4}
\psline[liftpen=1,linearc=\linearc](\xc,\ya)(\xc,\yb)(\xd,\yb)
\psline[liftpen=1,linearc=\linearc](\xd,\yc)(\xc,\yc)(\xc,\yd)
\psline[liftpen=1,linearc=\linearc](\xb,\yd)(\xb,\yc)(\xa,\yc)
\psline[liftpen=1,linearc=\linearc](\xa,\yb)(\xb,\yb)(\xb,\ya)}
\pscustom{%
\rotate{#4}
\psline[liftpen=1,linearc=\linearc](\xc,\ya)(\xc,\yb)(\xd,\yb)
\psline[liftpen=2,linearc=\linearc](\xd,\yc)(\xc,\yc)(\xc,\yd)
\psline[liftpen=2,linearc=\linearc](\xb,\yd)(\xb,\yc)(\xa,\yc)
\psline[liftpen=2,linearc=\linearc](\xa,\yb)(\xb,\yb)(\xb,\ya)
}
}
}
\newcommand{\fourvertexdbltr}[4][white]{%
\setlength{\xa}{0\unit}
\addtolength{\xa}{-1\unit}
\setlength{\xb}{0\unit}
\addtolength{\xb}{-0.5\dlinewidth}
\setlength{\xc}{0\unit}
\addtolength{\xc}{0.5\dlinewidth}
\setlength{\xd}{0\unit}
\addtolength{\xd}{1\unit}
\setlength{\ya}{0\unit}
\addtolength{\ya}{-1\unit}
\setlength{\yb}{0\unit}
\addtolength{\yb}{-0.5\dlinewidth}
\setlength{\yc}{0\unit}
\addtolength{\yc}{0.5\dlinewidth}
\setlength{\yd}{0\unit}
\addtolength{\yd}{1\unit}
\psset{doubleline=false}
\rput{0}(#2\unit,#3\unit){%
\pscustom[fillstyle=\sfillstyle,fillcolor=#1,linecolor=#1,linewidth=0pt]{%
\rotate{#4}
\psline[liftpen=1,linearc=1.5\linearc](\xc,\ya)(0,0)(\xa,\yc)
\psline[liftpen=1,linearc=\linearc](\xd,\yc)(\xc,\yc)(\xc,\yd)
\psline[liftpen=1,linearc=1.5\linearc](\xb,\yd)(0,0)(\xa,\yc)
\psline[liftpen=1,linearc=\linearc](\xa,\yb)(\xb,\yb)(\xb,\ya)}
\pscustom{%
\rotate{#4}
\psline[liftpen=1,linearc=1.5\linearc](\xc,\ya)(0,0)(\xa,\yc)
\psline[liftpen=2,linearc=\linearc](\xd,\yc)(\xc,\yc)(\xc,\yd)
\psline[liftpen=2,linearc=1.5\linearc](\xb,\yd)(0,0)(\xd,\yb)
\psline[liftpen=2,linearc=\linearc](\xa,\yb)(\xb,\yb)(\xb,\ya)
}
}
}

\newlength{\armlen}
\newcounter{nnodenum}
\newcounter{mnodenum}
\numberwithin{equation}{section}
\newcommand{\cvertr}[4][]{%
\settoheight{\eqoff}{$\times$}%
\setlength{\eqoff}{0.5\eqoff}%
\addtolength{\eqoff}{-9\unitlength}%
\raisebox{\eqoff}{%
\fmfframe(5,3)(5,3){%
\begin{fmfchar*}(12,12)
\fmfleft{v3,v1}
\fmfright{v2}
\fmf{#2}{v1,vc1}
\fmf{#3}{v2,vc1}
\fmf{#4}{v3,vc1}
\fmffreeze
\fmfposition
\fmfipath{p[]}
\fmfipair{vm[],vo[],vi[]}
\fmfiset{p1}{vpath(__v1,__vc1)}
\fmfiset{p2}{vpath(__v2,__vc1)}
\fmfiset{p3}{vpath(__v3,__vc1)}
\svertex{vm1}{p1}
\dvertex{vo1}{vi1}{p1}
\svertex{vm2}{p2}
\dvertex{vi2}{vo2}{p2}
\svertex{vm3}{p3}
\dvertex{vo3}{vi3}{p3}
{#1}
\end{fmfchar*}}}
}

\newcommand{\qvertr}[5][]{%
\settoheight{\eqoff}{$\times$}%
\setlength{\eqoff}{0.5\eqoff}%
\addtolength{\eqoff}{-12\unitlength}%
\raisebox{\eqoff}{%
\fmfframe(4,6)(4,6){%
\begin{fmfchar*}(12,12)
\fmfleft{v1,v4}
\fmfright{v2,v3}
\fmfforce{(0,h)}{v1}
\fmfforce{(0,0)}{v4}
\fmfforce{(w,0)}{v3}
\fmfforce{(w,h)}{v2}
\fmf{#2}{v1,vc1}
\fmf{#3}{v2,vc1}
\fmf{#4}{v3,vc1}
\fmf{#5}{v4,vc1}
\fmffreeze
\fmfposition
\fmfipath{p[]}
\fmfipair{vm[],vo[],vi[]}
\fmfiset{p1}{vpath(__v1,__vc1)}
\fmfiset{p2}{vpath(__v2,__vc1)}
\fmfiset{p3}{vpath(__v3,__vc1)}
\fmfiset{p4}{vpath(__v4,__vc1)}
\svertex{vm1}{p1}
\dvertex{vo1}{vi1}{p1}
\svertex{vm2}{p2}
\dvertex{vo2}{vi2}{p2}
\svertex{vm3}{p3}
\dvertex{vo3}{vi3}{p3}
\svertex{vm4}{p4}
\dvertex{vo4}{vi4}{p4}
{#1}
\end{fmfchar*}}}
}

\newcommand{\vacpolp}[2][0.5]{%
\fmfcmd{
begingroup;
save t,v,tv,do,di,ppol,pstr,dia;
path ppol,pstr;
pair v[],tv[],do[],di[];
ppol=#2;
t3=arctime (0.5*arclength ppol) of ppol;
v3=point t3 of ppol;
dia=#1 arclength ppol; 
fill(fullcircle scaled dia shifted v3) withcolor 0.2black;
endgroup;
}
}
\newcommand{\swfone}[4][]{%
\settoheight{\eqoff}{$\times$}%
\setlength{\eqoff}{0.5\eqoff}%
\addtolength{\eqoff}{-7.5\unitlength}%
\raisebox{\eqoff}{%
\fmfframe(4,0)(4,0){%
\begin{fmfchar*}(20,15)
\fmfleft{v1}
\fmfright{v2}
\fmffixed{(0.5w,0)}{vc1,vc2}
\fmf{#2}{v1,vc1}
\fmf{#2}{vc2,v2}
\fmf{#3}{vc1,vc2}
\fmf{#4}{vc2,vc1}
\fmffreeze
\fmfposition
\fmfipath{p[]}
\fmfipair{vm[]}
\fmfiset{p1}{vpath(__v1,__vc1)}
\fmfiset{p2}{vpath(__vc1,__vc2)}
\fmfiset{p3}{reverse vpath(__vc2,__vc1) }
\fmfiset{p4}{vpath(__vc2,__v2)}
\svertex{vm1}{p2}
\svertex{vm2}{p3}
{#1}
\end{fmfchar*}}}}

\begin{document}
\begin{fmffile}{graphs1}
\fmfcmd{%
thin := 1pt; 
thick := 2thin;
arrow_len := 4mm;
arrow_ang := 15;
curly_len := 3mm;
dash_len := 1.5mm; 
dot_len := 1mm; 
wiggly_len := 2mm; 
wiggly_slope := 60;
zigzag_len := 2mm;
zigzag_width := 2thick;
decor_size := 5mm;
dot_size := 2thick;
}

\fmfcmd{%
}

\fmfcmd{%
marksize=2mm;
def draw_mark(expr p,a) =
  begingroup
    save t,tip,dma,dmb; pair tip,dma,dmb;
    t=arctime a of p;
    tip =marksize*unitvector direction t of p;
    dma =marksize*unitvector direction t of p rotated -45;
    dmb =marksize*unitvector direction t of p rotated 45;
    linejoin:=beveled;
    draw (-.5dma.. .5tip-- -.5dmb) shifted point t of p;
  endgroup
enddef;
style_def derplain expr p =
    save amid;
    amid=.5*arclength p;
    draw_mark(p, amid);
    draw p;
enddef;
def draw_marks(expr p,a) =
  begingroup
    save t,tip,dma,dmb,dmo; pair tip,dma,dmb,dmo;
    t=arctime a of p;
    tip =marksize*unitvector direction t of p;
    dma =marksize*unitvector direction t of p rotated -45;
    dmb =marksize*unitvector direction t of p rotated 45;
    dmo =marksize*unitvector direction t of p rotated 90;
    linejoin:=beveled;
    draw (-.5dma.. .5tip-- -.5dmb) shifted point t of p withcolor 0white;
    draw (-.5dmo.. .5dmo) shifted point t of p;
  endgroup
enddef;
style_def derplains expr p =
    save amid;
    amid=.5*arclength p;
    draw_marks(p, amid);
    draw p;
enddef;
def draw_markss(expr p,a) =
  begingroup
    save t,tip,dma,dmb,dmo; pair tip,dma,dmb,dmo;
    t=arctime a of p;
    tip =marksize*unitvector direction t of p;
    dma =marksize*unitvector direction t of p rotated -45;
    dmb =marksize*unitvector direction t of p rotated 45;
    dmo =marksize*unitvector direction t of p rotated 90;
    linejoin:=beveled;
    draw (-.5dma.. .5tip-- -.5dmb) shifted point t of p withcolor 0white;
    draw (-.5dmo.. .5dmo) shifted point arctime a+0.25 mm of p of p;
    draw (-.5dmo.. .5dmo) shifted point arctime a-0.25 mm of p of p;
  endgroup
enddef;
style_def derplainss expr p =
    save amid;
    amid=.5*arclength p;
    draw_markss(p, amid);
    draw p;
enddef;
style_def dblderplain expr p =
    save amidm;
    save amidp;
    amidm=.5*arclength p-0.75mm;
    amidp=.5*arclength p+0.75mm;
    draw_mark(p, amidm);
    draw_mark(p, amidp);
    draw p;
enddef;
style_def dblderplains expr p =
    save amidm;
    save amidp;
    amidm=.5*arclength p-0.75mm;
    amidp=.5*arclength p+0.75mm;
    draw_mark(p, amidm);
    draw_marks(p, amidp);
    draw p;
enddef;
style_def dblderplainss expr p =
    save amidm;
    save amidp;
    amidm=.5*arclength p-0.75mm;
    amidp=.5*arclength p+0.75mm;
    draw_mark(p, amidm);
    draw_markss(p, amidp);
    draw p;
enddef;
style_def dblderplainsss expr p =
    save amidm;
    save amidp;
    amidm=.5*arclength p-0.75mm;
    amidp=.5*arclength p+0.75mm;
    draw_marks(p, amidm);
    draw_markss(p, amidp);
    draw p;
enddef;
}

\fmfcmd{%
style_def plain_ar expr p =
  cdraw p;
  shrink (0.6);
  cfill (arrow p);
  endshrink;
enddef;
style_def plain_rar expr p =
  cdraw p; 
  shrink (0.6);
  cfill (arrow reverse(p));
  endshrink;
enddef;
style_def dashes_ar expr p =
  draw_dashes p;
  shrink (0.6);
  cfill (arrow p);
  endshrink;
enddef;
style_def dashes_rar expr p =
  draw_dashes p;
  shrink (0.6);
  cfill (arrow reverse(p));
  endshrink;
enddef;
style_def dots_ar expr p =
  draw_dots p;
  shrink (0.6);
  cfill (arrow p);
  endshrink;
enddef;
style_def dots_rar expr p =
  draw_dots p;
  shrink (0.6);
  cfill (arrow reverse(p));
  endshrink;
enddef;
}


\begingroup\parindent0pt
\begin{flushright}\footnotesize
\texttt{HU-MATH-2013-15}\\
\texttt{HU-EP-13/40}
\end{flushright}
\vspace*{2em}
\centering
\begingroup\LARGE
\bf
Non-conformality of $\gamma_i$-deformed \\
$\mathcal{N}=4$ SYM theory
\par\endgroup
\vspace{1.5em}
\begingroup\large
{\bf Jan Fokken,
Christoph Sieg,
Matthias Wilhelm}
\par\endgroup
\vspace{1em}
\begingroup\itshape
Institut f\"ur Mathematik und Institut f\"ur Physik\\
Humboldt-Universit\"at zu Berlin\\
IRIS Geb\"aude \\
Zum Gro\ss{}en Windkanal 6 \\
12489 Berlin
\par\endgroup
\vspace{1em}
\begingroup\ttfamily
fokken, csieg, mwilhelm@physik.hu-berlin.de \\
\par\endgroup
\vspace{1.5em}
\endgroup

\thispagestyle{empty}
\paragraph{Abstract.}
We show that the $\gamma_i$-deformation, which was proposed as candidate gauge theory for a non-supersymmetric three-parameter deformation of the AdS/CFT correspondence, is not conformally invariant due to a running double-trace coupling -- not even in the 't Hooft limit. Moreover, this non-conformality cannot be cured when we extend the theory by adding at tree-level arbitrary multi-trace couplings that obey certain minimal consistency requirements. Our findings suggest a possible connection between this breakdown of conformal invariance and a puzzling divergence recently encountered in the integrability-based descriptions of two-loop finite-size corrections for the single-trace operator of two identical chiral fields. We propose a test to clarify this.



\newpage


\section{Introduction and summary}

\label{sec:introduction}

\subsection{General setup}
\label{subsec:setup}

The $\AdS/\CFT$ correspondence \cite{Maldacena:1997re,Gubser:1998bc,Witten:1998qj} predicts dualities between certain string theories in anti-de Sitter (AdS) space and conformal field theories (CFTs). Its most prominent example concerns type $\twob$ string theory in $\AdS_5\times\text{S}^5$ with $N$ units of five-form flux and the four-dimensional maximally ($\mathcal{N}=4$) supersymmetric Yang-Mills (SYM) theory with gauge group $SU(N)$. It is most accessible in the 't Hooft limit \cite{'tHooft:1973jz}, where $N\to\infty$ and the Yang-Mills coupling constant $g_\YM\to0$ such that the 't Hooft coupling $\lambda=g_\YM^2 N$ is kept fixed: the string theory becomes free, and in the gauge theory non-planar vacuum diagrams are suppressed.\footnote{Subtleties for diagrams with external legs will be discussed below.}

By applying discrete orbifold projections \cite{Kachru:1998ys,Lawrence:1998ja} or continuous deformations \cite{Lunin:2005jy,Frolov:2005ty,Frolov:2005dj,Frolov:2005iq} to this setup, further examples for such dualities with fewer (super)symmetries have been constructed; see \cite{Zoubos:2010kh} for a review.

In \cite{Lunin:2005jy}, Lunin and Maldacena formulated a deformation of the maximally supersymmetric duality, introducing one complex deformation parameter. When restricted to a real parameter, the deformed string background can be obtained by applying a TsT transformation, i.e.\ a combination of a T-duality, a shift (s) of an angular variable and another T-duality, to the $\text{S}^5$ factor of the $\AdS_5\times\text{S}^5$ background. This breaks the isometry group $SO(6)$ of the $\text{S}^5$ to its $U(1)\times U(1)\times U(1)$ Cartan subgroup. One specific combination of the latter becomes the R-symmetry of the preserved simple ($\mathcal{N}=1$) supersymmetry. The gauge-theory dual has been identified as a particular case of the Leigh-Strassler deformations \cite{Leigh:1995ep} of $\mathcal{N}=4$ SYM theory. This theory is called the $\beta$-deformation, where $\beta$ refers to the single real deformation parameter.

In order to break also the remaining supersymmetry and hence obtain a non-super\-symmetric example of the AdS/CFT correspondence, Frolov \cite{Frolov:2005dj} generalized the above construction by applying three TsT transformations to the string background, each depending on an individual angular shift parameter $\gamma_i$, $i=1,2,3$. He proposed that the dual gauge theory should be given by the so-called $\gamma_i$-deformation of the $\mathcal{N}=4$ SYM theory. In the subsequent paper \cite{Frolov:2005iq}, Frolov, Roiban, and Tseytlin made the gauge theory and the matching with the string theory more explicit. In the special case where all parameters assume a common value $\gamma_i=-\pi\beta$, $i=1,2,3$, $\mathcal{N}=1$ supersymmetry is restored and the $\beta$-deformation is recovered.

Both deformed gauge theories can be formulated using a non-commutative $\ast$-product that  introduces a phase depending on the three $U(1)\times U(1)\times U(1)$ Cartan charges of the respective fields.

\subsection{Conformal invariance}
\label{subsec:conformal}

The $\AdS_5$ factor of the string background has $SO(2,4)$ as isometry group, which -- according to the AdS/CFT correspondence -- must also be present in the gauge theory. Since $SO(2,4)$ is the conformal group in four dimensions, the dual gauge theory should be a conformal field theory. In the maximally supersymmetric example, this is indeed the case: the classical $\mathcal{N}=4$ SYM theory is trivially conformal. Even more important, the conformal symmetry is preserved in the quantized theory. The coupling constant is not renormalized and the $\beta$-function hence vanishes exactly such that no scale is introduced by quantum corrections. In fact, the $\mathcal{N}=4$ SYM theory is finite \cite{Mandelstam:1982cb,Brink:1982wv}, i.e.\ its observables are free of divergences.\footnote{Divergences do occur if gauge invariant composite operators are introduced as external states.} The aforementioned orbi\-fold projections and TsT transformations only act in the $\text{S}^5$ directions, keeping the $\AdS_5$ factor and thus its $SO(2,4)$ isometry group intact.
Therefore, the respective dual orbi\-fold gauge theories as well as the $\beta$- and $\gamma_i$-deformation of $\mathcal{N}=4$ SYM theory should be conformal field theories as well, at least if the resulting string background is stable and the $\AdS_5$-factor is exact.

The statement of conformal invariance is intimately related to the vanishing of the $\beta$-functions of all couplings, and hence we need to have a closer look at the structure of the couplings. Recall that the $\mathcal{N}=4$ SYM action only contains interactions in which the representation matrices of the gauge algebra appear in commutators and the contractions of their indices form a single trace. The $U(1)$ component of a field transforming under the $U(N)$ gauge group decouples from these commutator interactions, and hence the theories with $SU(N)$ and $U(N)$ gauge group are essentially the same.

When orbifolds or deformations are applied, these single-trace contributions transform into respective new single-trace terms. Moreover, new multi-trace couplings can occur. They are constructed from the twisted sectors of the orbifolds \cite{Dymarsky:2005uh} or $\ast$-deformed commutators. The latter are no longer antisymmetric under an exchange of their arguments and therefore distinguish between $SU(N)$ and $U(N)$ gauge groups \cite{Frolov:2005iq}. This difference between the gauge groups manifests itself e.g.\ for single-trace operators of two different chiral or anti-chiral scalars in the $\beta$-deformation: for $SU(N)$ and $U(N)$, they respectively have vanishing and non-vanishing one-loop anomalous dimensions \cite{Freedman:2005cg}. More importantly, quantum corrections involving the single-trace couplings may induce counter terms for double- and even higher multi-trace couplings in the $SU(N)$ and $U(N)$ theories, respectively. The emergence of such counter terms demands that the respective couplings are considered already at tree level. Indeed, for a $\mathds{Z}_2$ orbifold projection, one-loop contributions to double-trace couplings were found in \cite{Tseytlin:1999ii}.

The coefficients of the multi-trace couplings are subleading in $N$. Hence, in the 't Hooft limit, there is no backreaction to the original cubic and quartic single-trace terms in the action. In case of the orbifold projections, it was shown in \cite{Bershadsky:1998mb,Bershadsky:1998cb} that the properties of the single-trace terms are inherited from the parent $\mathcal{N}=4$ SYM theory. In case of the $\beta$- and $\gamma_i$-deformation, the inheritance of the finiteness of the single-trace couplings in the 't Hooft limit was proven respectively in \cite{Ananth:2006ac} and \cite{Ananth:2007px}. The argumentation closely follows the proof of finiteness of the $\mathcal{N}=4$ SYM theory \cite{Mandelstam:1982cb,Brink:1982wv}. One is hence tempted to draw the conclusion that the respective theories are conformal, at least in the 't Hooft limit, where the multi-trace couplings appear to be negligible. This conclusion is, however, premature. The decision whether a diagram contributes in the 't Hooft limit can a priori only be made for diagrams in which all color lines are closed, i.e.\ external lines have to be connected to external states. This subtlety already occurred in the context of finite-size (wrapping) corrections, and was analyzed in detail in \cite{Sieg:2005kd}. In the notation of \cite{Sieg:2005kd}, a diagram with external legs and without external states (composite operators) is called planar if it contributes at leading order in the $\frac{1}{N}$ expansion after a color-ordered contraction of its external legs with a single-trace vertex. Besides these diagrams, in the 't Hooft limit there may be contributions from non-planar diagrams, which effectively are multi-trace interactions.\footnote{Note that the propagators in the $SU(N)$ theory themselves contain double-trace terms. This will be discussed in detail in our upcoming publication \cite{Fokken:2013mza}.} The reason for this is that in diagrams with multi-trace couplings the $N$-power is enhanced if one of the traces in the product is fully contracted with a trace of the same length in another coupling or external state, i.e.\ gauge invariant composite operator. In this way, the multi-trace couplings can contribute at the leading (planar) order in $\frac{1}{N}$, even if their coefficients are of lower power in $N$ compared to the ones of the single-trace couplings. Therefore, the 't Hooft limit is sensitive to the seemingly suppressed multi-trace couplings. Since their properties are not inherited from the parent theory \cite{Bershadsky:1998mb,Bershadsky:1998cb}, they may have non-vanishing $\beta$-functions, implying the breakdown of conformal invariance \cite{Adams:2001jb} -- also in the 't Hooft limit.\footnote{In a different context, concerns about the occurrence of multi-trace terms have been expressed earlier in \cite{Csaki:1999uy}.} It is hence very important to extend the analysis of conformal invariance to the induced multi-trace couplings.

For orbifold theories, the $\beta$-functions of induced double-trace couplings were analyzed at one loop in \cite{Dymarsky:2005uh}. If the orbifold projections preserve some supersymmetry, the $\beta$-functions may have fix points that are functions of the 't Hooft coupling constant, defining a fix line passing through the origin of the coupling-constant space \cite{Dymarsky:2005uh}. In contrast to this, for the non-supersymmetric orbifold projections no example was found in which all $\beta$-functions have fix points \cite{Dymarsky:2005uh}. These findings amounted to a no-go theorem that no non-supersymmetric orbifold exists with such a perturbatively accessible fix line \cite{Dymarsky:2005nc}. Isolated Banks-Zaks fix points \cite{Banks:1981nn} might still exist, i.e.\ the two-loop corrections to the $\beta$-functions might cancel the one-loop contributions at a perturbative real value of $g_\YM$. The running of the double-trace couplings was related to the emergence of tachyons in the twisted sectors of the string theory \cite{Dymarsky:2005nc}, similar to earlier relations in the context of non-commutative field theories in \cite{Armoni:2003va}.\footnote{Unoriented non-supersymmetric string theories, such as type $0\,$B theory in certain orientifold projections, can be free of tachyons, see e.g.\ \cite{Blumenhagen:1999ns, Blumenhagen:1999uy,Angelantonj:1999qg}. In fact, a certain low-energy gauge-theory description was found to be free of running double-trace couplings \cite{Liendo:2011da}. We thank Adi Armoni for pointing this out.} The running of the double-trace couplings is also connected to dynamical symmetry breaking \cite{Pomoni:2008de}.

The occurrence of subleading double-trace couplings in the orbifold examples rises the question whether similar terms are also generated in the $\beta$- and $\gamma_i$-deformation, as posed earlier in \cite{Dymarsky:2005nc}. If at least one of the corresponding coupling constants is running without a fix point, conformal invariance is broken. Note that the renormalization of such couplings is not captured by the proofs in \cite{Ananth:2006ac} and \cite{Ananth:2007px} of all-order finiteness. These proofs only consider planar diagrams without external states and thus only single-trace couplings; they neglect non-planar diagrams -- in particular those also contributing in the 't Hooft limit. Furthermore, the applied prescription of replacing ordinary products by $\ast$-products is only well defined inside of color traces with vanishing net $U(1)\times U(1)\times U(1)$ charge.\footnote{For $i\neq j$, $\tr(\phi^i\phi^j)=\tr(\phi^j\phi^i)$ but $\tr(\phi^i\ast\phi^j)\neq\tr(\phi^j\ast\phi^i)$.} In particular, it cannot be applied to multi-trace couplings with charged individual trace factors. These are the couplings which are not captured by the non-planar inheritance principle formulated in \cite{Jin:2012np}.

At least in the supersymmetric $\beta$-deformed case with gauge group $SU(N)$, there are no running double-trace couplings\footnote{For at most quartic interactions, triple- or  quadruple-trace couplings cannot occur if the gauge-group generators are traceless.} induced, and hence the theory is conformal. This follows immediately from the fact that the theory is a special case of the conformal Leigh-Strassler deformations \cite{Leigh:1995ep}. Note that a non-vanishing F-term-type double-trace coupling is present, but it has a vanishing $\beta$-function. This double-trace term appears when the component action is derived from the $\beta$-deformed $\mathcal{N}=1$ superfield action: it is generated when the auxiliary F-term fields of the superpotential are integrated out. If instead one considers a $U(N)$ gauge group in the deformed theory and then integrates out the auxiliary fields, the double-trace coupling is absent at tree level. However, the coupling to the $U(1)$ field components is irrelevant, and hence the theory flows to the $SU(N)$ theory in the IR \cite{Hollowood:2004ek}, making only the $SU(N)$ theory conformal.

Obviously, the supersymmetry-based arguments of \cite{Leigh:1995ep} that guarantee conformal invariance cannot be applied to the non-supersymmetric $\gamma_i$-deformation of the Frolov setup. Hence, one has to explicitly check whether the $\beta$-functions of all multi-trace couplings that are required for quantum consistency identically vanish or at least have (perturbatively accessible) fix points as functions of the 't Hooft coupling constant, i.e.\ a fix line.

\subsection{Our setup and conclusions}
\label{subsec:conclusions}

In this paper, we find that the $\gamma_i$-deformation is not conformally invariant by identifying a double-trace coupling with non-vanishing $\beta$-function. Even if we generalize the $\gamma_i$-deformation by adding additional (tree-level) multi-trace couplings and consider either $SU(N)$ or $U(N)$ gauge group, this $\beta$-function  cannot be forced to vanish.

Our setup consists of the original $\gamma_i$-deformed action as proposed in \cite{Frolov:2005dj}, either with $SU(N)$ or $U(N)$ as gauge group, but supplemented by a priori arbitrary multi-trace couplings that obey the following requirements:
\begin{enumerate}
\item \label{renormalizability}
renormalizability by power counting,
\item \label{existplanar}
existence of the 't Hooft limit (no proliferation of $N$-power beyond
the planar order),
\item \label{U1preserve}
preservation of the three global $U(1)$ charges,
\item \label{betalimit} reduction to the supersymmetric $\beta$-deformation
in the special case $\gamma_1=\gamma_2=\gamma_3$.
\end{enumerate}
We also want to avoid that the differences between the $\gamma_i$-deformation and the $\beta$-deformation are postponed to the next loop order. Hence, we demand that at least one difference $\gamma_i-\gamma_j$, $i\neq j$, of two of the deformation angles must not be of the order of the effective planar coupling constant
\begin{equation}\label{coupldef}
g=\frac{\sqrt{\lambda}}{4\pi}\col\qquad \lambda=g_\YM^2N\pnt
\end{equation}

In this setup, we investigate the one-loop corrections to the multi-trace couplings. Their renormalizations receive contributions from the UV divergent one-particle irreducible (1PI) vertex corrections and wave function renormalization. Setting the respective combinations to zero yields the conditions for vanishing one-loop $\beta$-functions and hence for conformal invariance. These conditions form a system of coupled equations that are non-linear in the coupling tensors. We identify a particular component of the double-trace coupling given in \eqref{zombiekiller}, and reading e.g.\ for $i=1$
\begin{equation}
-\frac{g_\YM^2}{N}Q^{11}_{\text{F}\,11}\tr(\bar\phi_1\bar\phi_1)\tr(\phi^1\phi^1)
\col
\end{equation}
for which the respective equation in the system cannot be solved, i.e.\ which has the non-vanishing one-loop $\beta$-function given in \eqref{betaQres}. With the rearranged Yang-Mills coupling \eqref{coupldef}, its one-loop $\beta$-function for $U(N)$ as well as $SU(N)$ gauge group is
\begin{equation}
\begin{aligned}\label{betaQ1111res}
\beta_{Q_{\text{F}\,11}^{11}}
=4g^2\big((\cos\gamma_2-\cos\gamma_3)^2
+(Q^{11}_{\text{F}\,11})^2\big)
\pnt
\end{aligned}
\end{equation}

The expression in \eqref{betaQ1111res} is non-vanishing for generic $\gamma_i$ and any choice of the real tree-level double-trace coupling $Q^{11}_{\text{F}\,11}$. Hence, the $\gamma_i$-deformed theory, even if extended by multi-trace couplings, is not conformal -- not even in the `t Hooft limit. This leads to the following possibilities compatible with the AdS/CFT correspondence:
\begin{enumerate}
\item The string background is not stable because of the emergence of closed string tachyons. These should be related to the running multi-trace couplings, as was found in the non-supersymmetric orbifold setups in \cite{Dymarsky:2005nc}. In $\gamma_i$-deformed flat space, tachyons were found in \cite{Spradlin:2005sv}, but a clean connection with the instabilities of the $\gamma_i$-deformation has not yet been established.\footnote{We thank Radu Roiban for this comment.}
\item 
The Frolov background \cite{Frolov:2005dj} receives string corrections that deform the $\AdS_5$ part such that the $SO(2,4)$ symmetry is broken and hence the dual gauge theory is not a conformal field theory. If the $\AdS_5$ factor of the Frolov background should turn out to be exact, it might also be possible that the gauge theory dual to the Frolov background is not yet found. However, our results exclude all natural candidates and it may be that the dual CFT does not even have a Lagrangian description with the field content of $\mathcal{N}=4$ SYM theory. 
\item The deformation angles are functions of the 't Hooft coupling and agree at zero coupling, $\gamma_i-\gamma_j=\mathcal{O}(g)$. This is reminiscent of the situation in the ABJM and ABJ correspondences \cite{Aharony:2008ug,Aharony:2008gk} and in the interpolating quiver gauge theory of \cite{Gadde:2009dj}, where finite functions of the couplings were respectively found in \cite{Minahan:2009aq,Minahan:2009wg,Leoni:2010tb} and \cite{Pomoni:2011jj}. It is hard to exclude this possibility, since by adjusting the deformation angles the non-vanishing of the $\beta$-function \eqref{betaQ1111res} can always be postponed to the next order.
\end{enumerate}
It is of high importance to determine which of these possible outcomes is correct.\footnote{It would also be interesting to extend the calculation to two loops and investigate whether the $\gamma_i$-deformed theory exhibits Banks-Zaks fix points \cite{Banks:1981nn}. Note that besides \eqref{betaQ1111res} there exist two analogous $\beta$-functions for $i=2,3$ each of which depends on different angles $\gamma_i$. Hence, due to the three conditions for their vanishing, the form of a possible fix point in $g$ is highly restricted.} To this end, it would be particularly interesting to compute the one-loop corrections to the string background.\footnote{
The analysis based on D-instantons in \cite{Ferrari:2013pq} cannot capture the 
breakdown of conformal invariance, neither in the $\gamma_i$-deformation nor in the
$U(N)$ $\beta$-deformation. First, the double-trace contributions are formally
suppressed in $\frac{1}{N}$ and seem to be discarded by the formalism.
Second, the metric and dilaton-axion can be constructed from instanton probes to quadratic order \cite{Durnford:2006nb}, but the
full geometry can only be determined
to linear order \cite{Ferrari:2013pq} in the deformation
parameters $\gamma_i$. The breakdown of conformal invariance starts,
however, at quadratic order in the $\gamma_i$, as seen from the
$\beta$-function \eqref{betaQ1111res}.}

\subsection{Integrability}
\label{subsec:integrability}

An important consequence of the conformal symmetry in the gauge theory is that the (anomalous) scaling dimensions of gauge invariant composite operators become observables: since the $\beta$-functions of all couplings are zero, the anomalous dimensions are renormalization-scheme independent and can be measured as eigenvalues of the generator of dilatations, known as the dilatation operator. The AdS/CFT correspondence predicts that these scaling dimensions should match with the energies of respective string states in the gravitational theory. This has been a direction of intense studies in the last decade. In particular, in the 't Hooft limit the eigenvalue-problem shows signs of integrability, and this has led to enormous progress in testing and understanding the AdS/CFT correspondence, see the review collection \cite{Beisert:2010jr} for a comprehensive list of references. Single-trace operators of length $L$, i.e.\ those containing $L$ elementary fields, are mapped to cyclic spin chains of the same length. The dilatation operator is identified with the integrable Hamiltonian acting on these chains. Integrability was found not only in the original correspondence involving the $\mathcal{N}=4$ SYM theory, but also for the orbifold constructions and the Lunin-Maldacena and Frolov setups -- see \cite{Zoubos:2010kh} for a review. In the deformed theories, the claimed integrability can in particular be tested for single-impurity operators, which are not protected for length $L\geq 3$. They consist of $L-1$ chiral scalar fields of one flavor and a single one of another flavor.\footnote{Corresponding operators exist also in the orbifold theories \cite{Mukhi:2002ck,Wang:2003cu,Ideguchi:2004wm,DeRisi:2004bc,Beisert:2005he}.} Such single-impurity operators map to cyclic spin chains with a single magnon. This magnon has non-vanishing momentum and hence a non-vanishing energy, corresponding to a non-vanishing anomalous dimension of the whole state; this follows from the Bethe equations of the deformed theories \cite{Berenstein:2004ys,Frolov:2005ty,Beisert:2005if} into which the deformation enters via twisted boundary conditions.\footnote{The twisted Bethe ansatz can be derived from a twisted transfer matrix \cite{Gromov:2010dy} corresponding to operational twisted boundary conditions \cite{Arutyunov:2010gu} or, alternatively, a twisted S-matrix \cite{Ahn:2010ws}.}

In the supersymmetric $\beta$-deformation, the leading wrapping corrections to the anomalous dimensions of the aforementioned operators with three or more fields have been calculated in \cite{Fiamberti:2008sn}. In the integrability-based descriptions, these finite-size effects are captured in terms of L\"uscher corrections, Y-system and thermodynamic Bethe ansatz (TBA) -- see \cite{Janik:2010kd}, \cite{Bajnok:2010ke} and \cite{Gromov:2010kf} for reviews. By employing these descriptions, the results of \cite{Fiamberti:2008sn} were reproduced in \cite{Gunnesson:2009nn} for $\beta=\frac{1}{2}$ and in \cite{Gromov:2010dy} and \cite{Arutyunov:2010gu} for generic $\beta$. The work \cite{Arutyunov:2010gu} also provides higher-order wrapping corrections, also in various orbifold theories. For composite operators of two chiral scalar fields of different flavor, a logarithmic divergence was found in the leading finite-size correction. Such a divergence was encountered earlier in the expressions for the ground state energy of the TBA in the undeformed theory \cite{Frolov:2009in} and in non-supersymmetric orbifolds \cite{deLeeuw:2011rw}.

In the non-supersymmetric case, not only operators corresponding to single-magnon states acquire an anomalous dimension. Also the operators that are built from chiral (or anti-chiral) fields of a single flavor are no longer protected and acquire anomalous dimensions by finite-size effects, which were determined including also  double-wrapping corrections \cite{Ahn:2011xq}. Again, a logarithmic divergence was found for the first wrapping correction for an operator of two identical chiral scalar fields.

The meaning of the aforementioned divergences in the equations of TBA, L\"uscher and Y-system at $L=2$ is still unclear.\footnote{In \cite{deLeeuw:2012hp}, it was found that the divergent ground-state energy vanishes in the undeformed theory when a regulating twist is introduced in the $\AdS_5$ directions. This regularization extends to the ground state of the supersymmetric deformations. We thank Sergey Frolov for this comment.} Based on our observations, we will now describe a possible pathway for their investigation. As we have explained before, both, the $\beta$- and $\gamma_i$-deformations, distinguish between $U(N)$ and $SU(N)$ gauge groups -- unlike the $\mathcal{N}=4$ SYM theory. In particular, the (induced) double-trace couplings with charged individual trace factors are sensitive to the choice between the two types of gauge groups. This is most striking in the $\beta$-deformed case, where the F-term-type double-trace coupling breaks or preserves the conformal invariance in the $U(N)$ or $SU(N)$ case, respectively. In Subsection \ref{subsec:conformal}, we have argued that composite $L=2$ operators receive quantum corrections from these double-trace couplings in the 't Hooft limit. Among the ones composed only of chiral scalar fields, the $L=2$ operators are in fact the only ones receiving such corrections.\footnote{We will present a test of the leading finite-size corrections in the $\gamma_i$-deformation in \cite{Fokken:2014soa}.} In order to also describe the $L=2$ operators of the $SU(N)$ theories, one should modify the integrability-based TBA, L\"uscher and Y-system equations. In the $\beta$-deformation, this modification should remove the divergence at $L=2$, while not affecting any other results for the operators composed of two types of chiral scalars. Then, the analogous procedure should be applied in the $\gamma_i$-deformation. If the divergences are removed also there, it seems reasonable to identify the missing incorporation of this new finite-size effect\footnote{This effect, which we call prewrapping, is caused by double-trace couplings as will be explained in the upcoming work \cite{Fokken:2013mza}.} as the origin of the divergences. If, however, a divergence persists in the $\gamma_i$-deformed case, this suggests that the divergences are associated with the breakdown of conformal invariance in both the $U(N)$ $\beta$-deformation and $U(N)$ as well as $SU(N)$ $\gamma_i$-deformations. Note that the correct integrability-based descriptions must reproduce the vanishing anomalous dimensions for the operators with $L=2$ chiral scalar fields in the $\beta$-deformation with $SU(N)$ gauge group. In the $\gamma_i$-deformation with $SU(N)$ gauge group, however, even if the divergences are removed, the results need not match the finite field theory results\footnote{We will present the anomalous dimensions for operators of $L=2$ identical chiral scalars in \cite{Fokken:2014soa}.}. This possibility arises since the anomalous dimensions become scheme-dependent beyond one loop due to the breaking of conformal invariance. One might hence have to engineer a matching for the finite parts in order to fix a scheme, and then test this scheme choice by comparing with further data coming e.g.\ from other types of composite operators in the theory.

\subsection{Organization of this paper}

In Section \ref{sec:doubletrcoupl}, we start our analysis with the presentation of a brief argument that double-trace couplings in the $SU(N)$ $\beta$-deformation are already present at tree-level. We then introduce in Section \ref{sec:multitrdef} such couplings for the $SU(N)$ $\gamma_i$-deformation, and also further multi-trace couplings for the respective $U(N)$ gauge theory, which obey the restrictions \ref{renormalizability}.-\ref{betalimit}.\ listed in Subsection \ref{subsec:conclusions}. In Section \ref{sec:zombiekiller}, we identify a particular set of double-trace couplings that acquire UV-divergent one-loop corrections and hence have non-vanishing $\beta$-functions -- implying the breakdown of conformal invariance  for generic deformation parameters $\gamma_i$, $i=1,2,3$. Several appendices contain the action (\ref{app:action}), the Feynman rules (\ref{app:frules}), and auxiliary results necessary for the calculation (\ref{app:tensorid}--\ref{app:oneloopse}) as well as a short derivation of the $\beta$-function (\ref{app:couplren}).

\section{\texorpdfstring{Double-trace couplings in the $\beta$-deformation}{Double-trace couplings in the beta-deformation}}
\label{sec:doubletrcoupl}

In this section, we will demonstrate that in the $\beta$-deformation
with $SU(N)$ gauge group double-trace couplings are already present at 
tree-level.\footnote{The action including this double-trace terms
can also be found in \cite{Jin:2012np}. We thank Radu Roiban for pointing 
this out.}
We start from the action in terms of 
$\mathcal{N}=1$ superfields, which are multiplied by a superspace 
$\star$-product containing the deformation.
Expanding the superfields and $\star$-product in components and 
integrating out the 
auxiliary fields generates a double-trace coupling in the $SU(N)$ case, 
where no $U(1)$ component is present.   

The part of the Euclidean action of the $\beta$-deformed theory that depends on the
$\mathcal{N}=1$ chiral and anti-chiral superfields $\Phi^i$ and
$\bar\Phi_i$ 
assumes the form 
\begin{equation}\label{N1Smatter}
S_{\text{matter}}=\int\de^4x\,\de^4\theta\,
\tr(\e^{-g_\YM V}\bar\Phi_i\e^{g_\YM V}\Phi^i)
+\int\de^4x\,\de^2\theta\,W
+\int\de^4x\,\de^2\bar\theta\,\bar W
\col
\end{equation}
where the superpotential is given by
\begin{equation}\label{superpot}
W=\frac{i}{3!}g_\YM\epsilon_{ijk}\tr\big(\Phi^i\comm[\star]{\Phi^j}{\Phi^k}\big)
\pnt
\end{equation}
It involves a non-commutative $\star$-product of two superfields $\Phi^j$ and $\Phi^k$. When the products are expanded in terms of the fermionic coordinates of superspace, the superfields expand in their respective component fields and the superspace $\star$-product introduces phase factors which follow from the definitions in Appendix \ref{app:action} by setting $\gamma_1=\gamma_2=\gamma_3=-\pi\beta$. These phase factors can be captured in terms of the component field $\ast$-product \eqref{astproddef}.

Fixing the supergauge to the Wess-Zumino gauge, the component expansion of the action in \eqref{N1Smatter} contains the following terms
\begin{equation}\label{Scompwithaux}
S=\int\de^4x\,\tr\Big[\dots
+\bar F_iF^i
+\dots
+\frac{i}{2}g_\YM\epsilon_{ijk}F^i\comm[\ast]{\phi^j}{\phi^k}
+\frac{i}{2}g_\YM\epsilon^{ijk}\bar F_i\comm[\ast]{\bar\phi_j}{\bar\phi_k}
+\dots
\Big]
\col
\end{equation}
which depends on the chiral auxiliary fields $F^i$, chiral scalar fields $\phi^i$ and their respective conjugates. The first term stems from the first term in \eqref{N1Smatter}, while the second and third term are generated by the superpotential \eqref{superpot} and its complex conjugate, respectively.

In the next step, we integrate out the auxiliary fields and 
obtain\footnote{The first line of this equation can be found in  
\cite{Freedman:2005cg}.}
\begin{equation}
\begin{aligned}\label{Sdtr}
S&=\int\de^4x\,\Big[\dots
+\frac{g_\YM^2}{4}\epsilon^{ijk}\epsilon_{ilr}\tr(T^a\comm[\ast]{\bar\phi_j}{\bar\phi_k})\tr(T^a\comm[\ast]{\phi^l}{\phi^r})
+\dots
\Big]\\
&=
\int\de^4x\,\Big[\dots
+\frac{g_\YM^2}{2}\Big(\tr(\comm[\ast]{\bar\phi_i}{\bar\phi_j}\comm[\ast]{\phi^i}{\phi^j})
-\frac{s}{N}\tr(\comm[\ast]{\bar\phi_i}{\bar\phi_j}\big)\tr\big(\comm[\ast]{\phi^i}{\phi^j})\Big)
+\dots
\Big]
\col
\end{aligned}
\end{equation}
where the adjoint index $a=s,\dots,N^2-1$ is summed over, starting from $s=1$ for $SU(N)$ and $s=0$ for $U(N)$ gauge group. In the second line, we have used the second  of the following relations for the gauge group generators $\T^a$:
\begin{equation} \label{TTprod}
\tr(\T^a\T^b)=\delta^{ab}\col\qquad
\sum_{a=s}^{N^2-1}(\T^a)^i{}_j(\T^a)^k{}_l=\delta^i_l\delta^k_j-\frac{s}{N}\delta^i_j\delta^k_l
\pnt
\end{equation}
The first term in \eqref{Sdtr} is the quartic F-term interaction of the component action. The second term is the double-trace term. It is present in the $SU(N)$ theory, while it is absent in the $U(N)$ theory, at least at tree level. This leads to a difference between the  $SU(N)$ and $U(N)$ theory: the one-loop anomalous dimensions of operators of two different chiral or anti-chiral scalar fields vanish in the $SU(N)$ theory and are non-zero in the $U(N)$ theory \cite{Freedman:2005cg}. The double-trace coupling vanishes in the $\mathcal{N}=4$ SYM theory, where the non-commutative $\ast$-product reduces to the ordinary matrix product. In this case, the antisymmetry of the commutator is restored and its trace vanishes.

\section{Multi-trace deformations}
\label{sec:multitrdef}

In the previous section, we have demonstrated that in a component expansion of the $\beta$-deformed $\mathcal{N}=4$ SYM action with $SU(N)$ gauge group, double-trace couplings are present already at tree-level. Moreover, quantum corrections may lead to UV-divergent multi-trace terms. In this case, one is forced to introduce counter terms for these multi-trace couplings and also add respective tree-level couplings. In this section, we present the possible tensor structures which obey the conditions \ref{renormalizability}.-\ref{betalimit}.\ formulated in Subsection \ref{subsec:conclusions}.

For gauge group $SU(N)$, where all generators are traceless, the only possible multi-trace structure is a product of two length-two traces, such that the respective terms in the action assume the form\footnote{Note that in our conventions $\e^{S}$ with $S$ given in Appendix \ref{app:action} occurs in the Euclidean path integral. All terms in $S$ and in particular the couplings \eqref{dtc} have a reversed sign compared to those in an action $\tilde S$ occurring as $\e^{-\tilde S}$. As usual, a negative $\beta$-function corresponds to asymptotic freedom. The factor $g_\YM^2$ is chosen for convenience in order to match the coupling dependence of the quartic single-trace interactions in the action \eqref{S}.}
\begin{equation}\label{dtc}
-\frac{g_\YM^2}{N}\big[Q^{ij}_{\text{F}\,kl}\tr(\bar\phi_i\bar\phi_j)\tr(\phi^k\phi^l)
+Q^{ij}_{\text{D}\,kl}\tr(\bar\phi_i\phi^k)\tr(\bar\phi_j\phi^l)\big]\pnt
\end{equation} 
The condition of a real action in Euclidean space imposes the relations
\begin{equation}\label{conjQFD}
\begin{aligned}
(Q^{ij}_{\text{F}\,kl})^\ast&=Q^{kl}_{\text{F}\,ij}\col\qquad
(Q^{ij}_{\text{D}\,kl})^\ast=Q^{kl}_{\text{D}\,ij}\pnt
\end{aligned}
\end{equation}

In the $U(N)$ case, the $U(1)$ generator is not traceless, and this allows
us to supplement the action with cubic as well as further quartic multi-trace 
couplings. The cubic Yukawa-type couplings can be written as
\begin{equation}
\begin{aligned}\label{cUNc}
&\frac{g_\YM}{N}\big[
(\rho_{\psi\,i})_{BA}\tr(\psi^{\alpha\,A})\tr(\phi^i\psi_\alpha^B)
+(\rho_{\phi\,i})_{BA}\tr(\phi^i)\tr(\psi^{\alpha\,B}\psi_\alpha^A)
\\
&\hphantom{\frac{g_\YM}{N}\big[}
+(\rho_{\bar\psi}^{\dagger\,i})^{BA}\tr(\bar\psi^{\dot\alpha}_A)\tr(\bar\phi_i\bar\psi_{\dot\alpha\,B})
+(\rho_{\bar\phi}^{\dagger\,i})^{BA}\tr(\bar\phi_i)\tr(\bar\psi^{\dot\alpha}_B\bar\psi_{\dot\alpha\,A})
\\
&\hphantom{\frac{g_\YM}{N}\big[}
+(\tilde\rho_{\bar\psi\,i})^{BA}\tr(\bar\psi^{\dot\alpha}_A)\tr(\phi^i\bar\psi_{\dot\alpha\,B})
+(\tilde\rho_{\bar\phi\,i})^{BA}\tr(\phi^i)\tr(\bar\psi^{\dot\alpha}_B\bar\psi_{\dot\alpha\,A})
\\
&\hphantom{\frac{g_\YM}{N}\big[}
+(\tilde\rho_\psi^{\dagger\,i})_{BA}\tr(\psi^{\alpha\,A})\tr(\bar\phi_i\psi_\alpha^B)
+(\tilde\rho_{\bar\phi}^{\dagger\,i})_{BA}\tr(\bar\phi_i)\tr(\psi^{\alpha\,B}\psi_\alpha^A)
\big]
\\
&+\frac{g_\YM}{N^2}\big[
(\rho_{3\,i})_{BA}\tr(\psi^{\alpha\,A})\tr(\phi^i)\tr(\psi_\alpha^B)
+(\rho_3^{\dagger\,i})^{BA}\tr(\bar\psi^{\dot\alpha}_A)\tr(\bar\phi_i)\tr(\bar\psi_{\dot\alpha\,B})\\
&\phantom{{}+{}\frac{g_\YM}{N^2}\big[}
+(\tilde\rho_{3\,i})^{BA}\tr(\bar\psi^{\dot\alpha}_A)\tr(\phi^i)\tr(\bar\psi_{\dot\alpha\,B})
+(\tilde\rho_3^{\dagger\,i})_{BA}\tr(\psi^{\alpha\,A})\tr(\bar\phi_i)\tr(\psi_\alpha^B)
\big]
\pnt
\end{aligned}
\end{equation}
Moreover, quartic interactions can be added, in which one or more traces with 
a single field occur. They read
\begin{equation}
\begin{aligned}\label{qUNc}
&-\frac{g_\YM^2}{N}\big[
Q^{ij}_{\bar\phi\,kl}\tr(\bar\phi_i)\tr(\bar\phi_j\phi^k\phi^l)
+Q^{ij}_{\phi\,kl}\tr(\phi^k)\tr(\bar\phi_i\bar\phi_j\phi^l)\big]\\
&-\frac{g_\YM^2}{N^2}\big[
Q^{ij}_{\bar\phi\bar\phi\,kl}\tr(\bar\phi_i)\tr(\bar\phi_j)\tr(\phi^k\phi^l)
+Q^{ij}_{\phi\phi\,kl}\tr(\phi^k)\tr(\phi^l)\tr(\bar\phi_i\bar\phi_j)\\
&\hphantom{{}-{}\frac{g_\YM^2}{N^2}\big[}+Q^{ij}_{\bar\phi\phi\,kl}\tr(\bar\phi_i)\tr(\phi^k)\tr(\bar\phi_j\phi^l)\big]\\
&-\frac{g_\YM^2}{N^3}
Q^{ij}_{4\,kl}\tr(\bar\phi_i)\tr(\bar\phi_j)\tr(\phi^k)\tr(\phi^l)
\pnt
\end{aligned}
\end{equation}
In the above combinations, we have explicitly separated all $U(1)$ fields from $SU(N)$ fields: each $U(1)$ component is written as a trace over the respective $U(N)$ field, whereas traces of more than one field are understood to contain only the $SU(N)$ components. The condition of a real action in Euclidean space imposes the following relations for the additional coupling tensors:
\begin{equation}\label{conjQ}
\begin{aligned}
(Q^{ij}_{\phi\,kl})^\ast&=Q^{kl}_{\bar\phi\,ji}\col\qquad
(Q^{ij}_{\bar\phi\phi\,kl})^\ast=Q^{kl}_{\bar\phi\phi\,ij}\col\qquad
(Q^{ij}_{\phi\phi\,kl})^\ast=Q^{kl}_{\bar\phi\bar\phi\,ij}\col\qquad
(Q^{ij}_{4\,kl})^\ast=Q^{kl}_{4\,ij}\pnt
\end{aligned}
\end{equation}

Note that the requirement \ref{existplanar}.\ of Subsection \ref{subsec:conclusions} restricts the $N$-powers of the multi-trace couplings: a coupling with $n$ traces must be suppressed by a factor of at least $N^{1-n}$ relative to a single-trace coupling, as shown in the following. Consider $n$ external single-trace states each of which can be planarly contracted with one of the individual single-trace factors in the $n$-trace coupling. In this contraction, the $n$ traces of the coupling generate a factor of $N^n$. If instead the external states are contracted in a single-trace coupling, only a single factor of $N$ is generated. Hence, the suppression factor $N^{1-n}$ is required to avoid a proliferation of $N$-power beyond the planar order.

Moreover, requirement 3.\ implies the vanishing of all tensor components of interactions that do not preserve the three global $U(1)$ charges.

\section{Running double-trace couplings}
\label{sec:zombiekiller}

In this section, we investigate the one-loop correction to a particular double-trace coupling that is contained in the F-term-type interaction of \eqref{dtc} and yields a vertex of four scalars with identical field flavor. The relevant terms that enter the action read
\begin{equation}\label{zombiekiller}
-\frac{g_\YM^2}{N}Q^{ii}_{\text{F}\,ii}\tr(\bar\phi_i\bar\phi_i)\tr(\phi^i\phi^i)
\col
\end{equation}
where $i$ assumes one of the three different values $i=1,2,3$ and throughout this section is never summed over. Below, we will first show that the couplings of these three individual terms are renormalized and hence running in the $SU(N)$ case. This cannot be avoided by extending the gauge group to $U(N)$, adding also the multi-trace terms  \eqref{cUNc}, \eqref{qUNc} to the action. For notational simplicity, we will abbreviate color traces with adjoint indices $a_1,\dots a_n=s,\dots,N^2-1$ as 
\begin{equation}\label{trdef}
\tr\big(\T^{a_1}\cdots\T^{a_n}\big)=\big(a_1\cdots a_n\big)
\pnt
\end{equation}

\subsection{\texorpdfstring{$SU(N)$ gauge group}{SU(N) gauge group}}
\label{subsec:SUN}

\begin{figure}[!ht]
\begin{center}
\subfigure[]{\label{ss4s4}%
\begin{pspicture}(-7.5,-7.5)(7.5,7.5)
\fourvertex{-3}{0}{45}
\fourvertex{3}{0}{45}
\setlength{\xa}{1.5\unit}
\addtolength{\xa}{-0.5\doublesep}
\addtolength{\xa}{-\linew}
\setlength{\xb}{4.5\unit}
\addtolength{\xb}{0.5\doublesep}
\addtolength{\xb}{\linew}
\setlength{\xc}{7.5\unit}
\addtolength{\xc}{-0.5\doublesep}
\addtolength{\xc}{-\linew}
\setlength{\xd}{10.5\unit}
\addtolength{\xd}{0.5\doublesep}
\addtolength{\xd}{\linew}
\setlength{\ya}{6\unit}
\addtolength{\ya}{0.5\doublesep}
\addtolength{\ya}{\linew}
\setlength{\yb}{3\unit}
\addtolength{\yb}{-0.5\doublesep}
\addtolength{\yb}{-\linew}
\psset{linecolor=black,doubleline=true}
\psbezier(-3.7071,0.7071)(-4.7071,1.7071)(-2.7071,2.7071)(0,0)
\psbezier(0,0)(2.7071,-2.7071)(4.7071,-1.7071)(3.7071,-0.7071)
\psset{linecolor=lightgray,doubleline=false}
\psbezier(-3.7071,0.7071)(-4.7071,1.7071)(-2.7071,2.7071)(0,0)
\psbezier(0,0)(2.7071,-2.7071)(4.7071,-1.7071)(3.7071,-0.7071)
\psset{linecolor=black,doubleline=true}
\psbezier(-2.2921,-0.7071)(-0.87868,-2.2071)(0.87868,2.12132)(2.2929,0.7071)
\psset{linecolor=lightgray,doubleline=false}
\psbezier(2.2929,0.7071)(0.87868,2.12132)(-0.87868,-2.2071)(-2.2921,-0.7071)
\psline(3.7071,-0.7071)(2.2929,0.7071)
\psline(-2.2921,-0.7071)(-3.7071,0.7071)
\psset{linecolor=black,doubleline=true}
\rput[B](-7,6.25){$a$}
\psbezier(-2.2921,0.7071)(-0.2921,2.7071)(-3,3)(-6,6)
\rput[B](7,6.25){$b$}
\psbezier(3.7071,0.7071)(5.7071,2.7071)(3,3)(6,6)
\rput[B](7,-7.125){$c$}
\psbezier(6,-6)(3,-3)(0.2921,-2.7071)(2.2921,-0.7071)
\rput[B](-7,-7.125){$d$}
\psbezier(-6,-6)(-3,-3)(-5.7071,-2.7071)(-3.7071,-0.7071)
\psset{linecolor=gray,doubleline=false}
\psbezier(-6,-6)(-3,-3)(-5.7071,-2.7071)(-3.7071,-0.7071)
\psline(-3.7071,-0.7071)(-2.2921,0.7071)
\psbezier(-2.2921,0.7071)(-0.2921,2.7071)(-3,3)(-6,6)
\psbezier(6,-6)(3,-3)(0.2921,-2.7071)(2.2921,-0.7071)
\psline(2.2921,-0.7071)(3.7071,0.7071)
\psbezier(3.7071,0.7071)(5.7071,2.7071)(3,3)(6,6)
\end{pspicture}}
\subfigure[]{\label{us4s4}%
\begin{pspicture}(-7.5,-7.5)(7.5,7.5)
\fourvertex{-3}{0}{45}
\fourvertex{3}{0}{45}
\setlength{\xa}{1.5\unit}
\addtolength{\xa}{-0.5\doublesep}
\addtolength{\xa}{-\linew}
\setlength{\xb}{4.5\unit}
\addtolength{\xb}{0.5\doublesep}
\addtolength{\xb}{\linew}
\setlength{\xc}{7.5\unit}
\addtolength{\xc}{-0.5\doublesep}
\addtolength{\xc}{-\linew}
\setlength{\xd}{10.5\unit}
\addtolength{\xd}{0.5\doublesep}
\addtolength{\xd}{\linew}
\setlength{\ya}{6\unit}
\addtolength{\ya}{0.5\doublesep}
\addtolength{\ya}{\linew}
\setlength{\yb}{3\unit}
\addtolength{\yb}{-0.5\doublesep}
\addtolength{\yb}{-\linew}
\psset{linecolor=black,doubleline=true}
\psbezier(-3.7071,0.7071)(-4.7071,1.7071)(-2.7071,2.7071)(0,0)
\psbezier(0,0)(2.7071,-2.7071)(4.7071,-1.7071)(3.7071,-0.7071)
\psset{linecolor=lightgray,doubleline=false}
\psbezier(-3.7071,0.7071)(-4.7071,1.7071)(-2.7071,2.7071)(0,0)
\psbezier(0,0)(2.7071,-2.7071)(4.7071,-1.7071)(3.7071,-0.7071)
\psset{linecolor=black,doubleline=true}
\psbezier(-2.2921,-0.7071)(-0.87868,-2.2071)(0.87868,2.12132)(2.2929,0.7071)
\psset{linecolor=lightgray,doubleline=false}
\psbezier(2.2929,0.7071)(0.87868,2.12132)(-0.87868,-2.2071)(-2.2921,-0.7071)
\psline(3.7071,-0.7071)(2.2929,0.7071)
\psline(-2.2921,-0.7071)(-3.7071,0.7071)
\psset{linecolor=black,doubleline=true}
\rput[B](-7,6.25){$a$}
\psbezier(-2.2921,0.7071)(-0.2921,2.7071)(-3,3)(-6,6)
\rput[B](7,6.25){$b$}
\psbezier(3.7071,0.7071)(5.7071,2.7071)(3,3)(6,6)
\rput[B](-7,-7.125){$d$}
\psbezier(-6,-6)(-3,-3)(0.2921,-2.7071)(2.2921,-0.7071)
\psset{linecolor=gray,doubleline=false}
\psbezier(-6,-6)(-3,-3)(0.2921,-2.7071)(2.2921,-0.7071)
\psline(2.2921,-0.7071)(3.7071,0.7071)
\psbezier(3.7071,0.7071)(5.7071,2.7071)(3,3)(6,6)
\psset{linecolor=black,doubleline=true}
\rput[B](7,-7.125){$c$}
\psbezier(6,-6)(3,-3)(-5.7071,-2.7071)(-3.7071,-0.7071)
\psset{linecolor=gray,doubleline=false}
\psbezier(6,-6)(3,-3)(-5.7071,-2.7071)(-3.7071,-0.7071)
\psline(-3.7071,-0.7071)(-2.2921,0.7071)
\psbezier(-2.2921,0.7071)(-0.2921,2.7071)(-3,3)(-6,6)
\end{pspicture}}
\subfigure[]{\label{ts4s4}%
\begin{pspicture}(-7.5,-7.5)(7.5,7.5)
\fourvertexdbltr{0}{-3}{45}
\fourvertexdbltr{0}{3}{45}
\setlength{\xa}{1.5\unit}
\addtolength{\xa}{-0.5\doublesep}
\addtolength{\xa}{-\linew}
\setlength{\xb}{4.5\unit}
\addtolength{\xb}{0.5\doublesep}
\addtolength{\xb}{\linew}
\setlength{\xc}{7.5\unit}
\addtolength{\xc}{-0.5\doublesep}
\addtolength{\xc}{-\linew}
\setlength{\xd}{10.5\unit}
\addtolength{\xd}{0.5\doublesep}
\addtolength{\xd}{\linew}
\setlength{\ya}{6\unit}
\addtolength{\ya}{0.5\doublesep}
\addtolength{\ya}{\linew}
\setlength{\yb}{3\unit}
\addtolength{\yb}{-0.5\doublesep}
\addtolength{\yb}{-\linew}
\psset{doubleline=true}
\psbezier(-0.7071,-2.2921)(-2.2071,-0.87868)(-2.2071,0.87868)(-0.7071,2.2929)
\psbezier(0.7071,2.2929)(2.2071,0.87868)(2.2071,-0.87868)(0.7071,-2.2921)
\rput[B](-7,6.25){$a$}
\psbezier(-0.7071,3.7071)(-2.7071,5.7071)(-3,3)(-6,6)
\rput[B](7,6.25){$b$}
\psbezier(0.7071,3.7071)(2.7071,5.7071)(3,3)(6,6)
\rput[B](7,-7.125){$c$}
\psbezier(6,-6)(3,-3)(2.7071,-5.7071)(0.7071,-3.7071)
\rput[B](-7,-7.125){$d$}
\psbezier(-0.7071,-3.7071)(-2.7071,-5.7071)(-3,-3)(-6,-6)
\psset{linecolor=gray,doubleline=false}
\psbezier(-6,-6)(-3,-3)(-2.7071,-5.7071)(-0.7071,-3.7071)
\psbezier(6,-6)(3,-3)(2.7071,-5.7071)(0.7071,-3.7071)
\psbezier(-0.7071,3.7071)(-2.7071,5.7071)(-3,3)(-6,6)
\psbezier(0.7071,3.7071)(2.7071,5.7071)(3,3)(6,6)
\psbezier(-0.7071,-2.2921)(-2.2071,-0.87868)(-2.2071,0.87868)(-0.7071,2.2929)
\psbezier(0.7071,-2.2921)(2.2071,-0.87868)(2.2071,0.87868)(0.7071,2.2929)
\end{pspicture}}
\subfigure[]{\label{tDs4s4}%
\begin{pspicture}(-7.5,-7.5)(7.5,7.5)
\fourvertexdbltr{0}{-3}{45}
\fourvertex{0}{3}{45}
\setlength{\xa}{1.5\unit}
\addtolength{\xa}{-0.5\doublesep}
\addtolength{\xa}{-\linew}
\setlength{\xb}{4.5\unit}
\addtolength{\xb}{0.5\doublesep}
\addtolength{\xb}{\linew}
\setlength{\xc}{7.5\unit}
\addtolength{\xc}{-0.5\doublesep}
\addtolength{\xc}{-\linew}
\setlength{\xd}{10.5\unit}
\addtolength{\xd}{0.5\doublesep}
\addtolength{\xd}{\linew}
\setlength{\ya}{6\unit}
\addtolength{\ya}{0.5\doublesep}
\addtolength{\ya}{\linew}
\setlength{\yb}{3\unit}
\addtolength{\yb}{-0.5\doublesep}
\addtolength{\yb}{-\linew}
\psset{doubleline=true}
\psbezier(-0.7071,-2.2921)(-2.2071,-0.87868)(-2.2071,0.87868)(-0.7071,2.2929)
\psbezier(0.7071,2.2929)(2.2071,0.87868)(2.2071,-0.87868)(0.7071,-2.2921)
\rput[B](-7,6.25){$a$}
\psbezier(-0.7071,3.7071)(-2.7071,5.7071)(-3,3)(-6,6)
\rput[B](7,6.25){$b$}
\psbezier(0.7071,3.7071)(2.7071,5.7071)(3,3)(6,6)
\rput[B](7,-7.125){$c$}
\psbezier(6,-6)(3,-3)(2.7071,-5.7071)(0.7071,-3.7071)
\rput[B](-7,-7.125){$d$}
\psbezier(-0.7071,-3.7071)(-2.7071,-5.7071)(-3,-3)(-6,-6)
\psset{linecolor=gray,doubleline=false}
\psbezier(-6,-6)(-3,-3)(-2.7071,-5.7071)(-0.7071,-3.7071)
\psline(-0.7071,2.2929)(0.7071,3.7071)
\psline(0.7071,2.2929)(-0.7071,3.7071)
\psbezier(6,-6)(3,-3)(2.7071,-5.7071)(0.7071,-3.7071)
\psbezier(-0.7071,3.7071)(-2.7071,5.7071)(-3,3)(-6,6)
\psbezier(0.7071,3.7071)(2.7071,5.7071)(3,3)(6,6)
\psbezier(-0.7071,-2.2921)(-2.2071,-0.87868)(-2.2071,0.87868)(-0.7071,2.2929)
\psbezier(0.7071,-2.2921)(2.2071,-0.87868)(2.2071,0.87868)(0.7071,2.2929)
\end{pspicture}}

\subfigure[]{\label{s4gg2}%
\begin{pspicture}(-7.5,-7.5)(7.5,7.5)
\fourvertexdbltr{0}{-3}{45}
\threevertex{-3}{2}{30}
\threevertex{3}{2}{150}
\setlength{\xa}{1.5\unit}
\addtolength{\xa}{-0.5\doublesep}
\addtolength{\xa}{-\linew}
\setlength{\xb}{4.5\unit}
\addtolength{\xb}{0.5\doublesep}
\addtolength{\xb}{\linew}
\setlength{\xc}{7.5\unit}
\addtolength{\xc}{-0.5\doublesep}
\addtolength{\xc}{-\linew}
\setlength{\xd}{10.5\unit}
\addtolength{\xd}{0.5\doublesep}
\addtolength{\xd}{\linew}
\setlength{\ya}{6\unit}
\addtolength{\ya}{0.5\doublesep}
\addtolength{\ya}{\linew}
\setlength{\yb}{3\unit}
\addtolength{\yb}{-0.5\doublesep}
\addtolength{\yb}{-\linew}
\psset{doubleline=true}
\psbezier(-2.134,2.5)(-1.268,3)(1.268,3)(2.134,2.5)
\rput[B](-7,6.25){$a$}
\psbezier(-0.7071,-2.2921)(-2.2071,-0.87868)(-2,0.268)(-2.5,1.134)
\psbezier(-3.5,2.866)(-4,3.732)(-4,4)(-6,6)
\rput[B](7,6.25){$b$}
\psbezier(0.7071,-2.2921)(2.2071,-0.87868)(2,0.268)(2.5,1.134)
\psbezier(3.5,2.866)(4,3.732)(4,4)(6,6)
\rput[B](7,-7.125){$c$}
\psbezier(6,-6)(3,-3)(2.7071,-5.7071)(0.7071,-3.7071)
\rput[B](-7,-7.125){$d$}
\psbezier(-0.7071,-3.7071)(-2.7071,-5.7071)(-3,-3)(-6,-6)
\psset{linecolor=gray,doubleline=false}
\psbezier(-6,-6)(-3,-3)(-2.7071,-5.7071)(-0.7071,-3.7071)
\psbezier(6,-6)(3,-3)(2.7071,-5.7071)(0.7071,-3.7071)
\psbezier(-0.7071,-2.2921)(-2.2071,-0.87868)(-2,0.268)(-2.5,1.134)
\psline(-2.5,1.134)(-3.5,2.866)
\psbezier(-3.5,2.866)(-4,3.732)(-4,4)(-6,6)
\psbezier(0.7071,-2.2921)(2.2071,-0.87868)(2,0.268)(2.5,1.134)
\psline(2.5,1.134)(3.5,2.866)
\psbezier(3.5,2.866)(4,3.732)(4,4)(6,6)
\end{pspicture}}
\subfigure[]{\label{s4gg3}%
\begin{pspicture}(-7.5,-7.5)(7.5,7.5)
\psset{doubleline=true}
\psbezier(-3.866,1.5)(-4.732,1)(-4.232,0.134)(-3.366,0.634)
\psbezier(-3.366,0.634)(-2.5,1.134)(2.5,1.134)(3.366,0.634)
\psbezier(3.866,1.5)(4.732,1)(4.232,0.134)(3.366,0.634)
\fourvertexdbltr{0}{-3}{45}
\threevertex{-3}{2}{210}
\threevertex{3}{2}{-30}
\setlength{\xa}{1.5\unit}
\addtolength{\xa}{-0.5\doublesep}
\addtolength{\xa}{-\linew}
\setlength{\xb}{4.5\unit}
\addtolength{\xb}{0.5\doublesep}
\addtolength{\xb}{\linew}
\setlength{\xc}{7.5\unit}
\addtolength{\xc}{-0.5\doublesep}
\addtolength{\xc}{-\linew}
\setlength{\xd}{10.5\unit}
\addtolength{\xd}{0.5\doublesep}
\addtolength{\xd}{\linew}
\setlength{\ya}{6\unit}
\addtolength{\ya}{0.5\doublesep}
\addtolength{\ya}{\linew}
\setlength{\yb}{3\unit}
\addtolength{\yb}{-0.5\doublesep}
\addtolength{\yb}{-\linew}
\psset{doubleline=true}
\rput[B](-7,6.25){$a$}
\psbezier(-0.7071,-2.2921)(-2.2071,-0.87868)(-2,0.268)(-2.5,1.134)
\psbezier(-3.5,2.866)(-4,3.732)(-4,4)(-6,6)
\rput[B](7,6.25){$b$}
\psbezier(0.7071,-2.2921)(2.2071,-0.87868)(2,0.268)(2.5,1.134)
\psbezier(3.5,2.866)(4,3.732)(4,4)(6,6)
\rput[B](7,-7.125){$c$}
\psbezier(6,-6)(3,-3)(2.7071,-5.7071)(0.7071,-3.7071)
\rput[B](-7,-7.125){$d$}
\psbezier(-0.7071,-3.7071)(-2.7071,-5.7071)(-3,-3)(-6,-6)
\psset{linecolor=gray,doubleline=false}
\psbezier(-6,-6)(-3,-3)(-2.7071,-5.7071)(-0.7071,-3.7071)
\psbezier(6,-6)(3,-3)(2.7071,-5.7071)(0.7071,-3.7071)
\psbezier(-0.7071,-2.2921)(-2.2071,-0.87868)(-2,0.268)(-2.5,1.134)
\psline(-2.5,1.134)(-3.5,2.866)
\psbezier(-3.5,2.866)(-4,3.732)(-4,4)(-6,6)
\psbezier(0.7071,-2.2921)(2.2071,-0.87868)(2,0.268)(2.5,1.134)
\psline(2.5,1.134)(3.5,2.866)
\psbezier(3.5,2.866)(4,3.732)(4,4)(6,6)
\end{pspicture}}

\subfigure[]{\label{fbox1}%
\begin{pspicture}(-8.5,-8.5)(8.5,8.5)
\recthreevertex{-3}{3}{135}
\recthreevertex{3}{3}{45}
\recthreevertexoneside{3}{-3}{-45}
\recthreevertexoneside{-3}{-3}{-135}
\psset{linecolor=black,doubleline=true}
\psline(-1.85,3)(1.85,3)
\psline(3,1.85)(3,-1.85)
\psline(1.85,-3)(-1.85,-3)
\psline(-3,-1.85)(-3,1.85)
\psset{linecolor=lightgray,doubleline=false,linestyle=dashed}
\psline(-3,1.85)(-3,3)
\psline(-3,3)(-1.85,3)
\psline(-1.85,3)(1.85,3)
\psline(1.85,3)(3,3)(3,1.85)
\psline(3,1.85)(3,-1.85)
\psline[linearc=\linearc](3,-1.85)(3,-3)(1.85,-3)
\psline(1.85,-3)(-1.85,-3)
\psline[linearc=\linearc](-1.85,-3)(-3,-3)(-3,-1.85)
\psline(-3,-1.85)(-3,1.85)

\psset{linecolor=black,doubleline=true,linestyle=solid}

\rput[B](-7,6.25){$a$}
\psline(-3.2,3.2)(-6,6)
\rput[B](7,6.25){$b$}
\psline(3.2,3.2)(6,6)
\rput[B](7,-7.125){$c$}
\psbezier(2.2,-2.2)(1.2,-1.2)(0.2,-2.2)(1.2,-3.2)
\psbezier(1.2,-3.2)(2.2,-4.2)(3,-3)(6,-6)
\rput[B](-7,-7.125){$d$}
\psbezier(-2.2,-2.2)(-1.2,-1.2)(-0.2,-2.2)(-1.2,-3.2)
\psbezier(-1.2,-3.2)(-2.2,-4.2)(-3,-3)(-6,-6)
\psset{linecolor=gray,doubleline=false}
\psline(-3,3)(-3.2,3.2)
\psline(-3.2,3.2)(-6,6)
\psline(3,3)(3.2,3.2)
\psline(3.2,3.2)(6,6)
\psline(2.8,-2.8)(2.2,-2.2)
\psbezier(2.2,-2.2)(1.2,-1.2)(0.2,-2.2)(1.2,-3.2)
\psbezier(1.2,-3.2)(2.2,-4.2)(3,-3)(6,-6)
\psline(-2.8,-2.8)(-2.2,-2.2)
\psbezier(-2.2,-2.2)(-1.2,-1.2)(-0.2,-2.2)(-1.2,-3.2)
\psbezier(-1.2,-3.2)(-2.2,-4.2)(-3,-3)(-6,-6)
\end{pspicture}}
\subfigure[]{\label{fbox2}%
\begin{pspicture}(-8.5,-8.5)(8.5,8.5)
\recthreevertex{-3}{3}{135}
\recthreevertex{3}{3}{45}
\recthreevertex{3}{-3}{-45}
\recthreevertex{-3}{-3}{-135}
\psset{linecolor=black,doubleline=true}
\psline(-1.85,3)(1.85,3)
\psbezier(3,1.85)(3,0)(-3,0)(-3,-1.85)
\psline(-1.85,-3)(1.85,-3)
\psset{linecolor=lightgray,doubleline=false,linestyle=dashed}
\psline(-3,1.85)(-3,3)
\psline(-3,3)(-1.85,3)
\psline(-1.85,3)(1.85,3)
\psline(1.85,3)(3,3)(3,1.85)
\psbezier(3,1.85)(3,0)(-3,0)(-3,-1.85)
\psline[linearc=\linearc](-3,-1.85)(-3,-3)(-1.85,-3)
\psline(-1.85,-3)(1.85,-3)
\psline[linearc=\linearc](1.85,-3)(3,-3)(3,-1.85)
\psset{linecolor=black,doubleline=true,linestyle=solid}
\psbezier(3,-1.85)(3,0)(-3,0)(-3,1.85)
\psset{linecolor=lightgray,doubleline=false,linestyle=dashed}
\psbezier(3,-1.85)(3,0)(-3,0)(-3,1.85)
\psset{linecolor=black,doubleline=true,linestyle=solid}

\rput[B](-7,6.25){$a$}
\psline(-3.2,3.2)(-6,6)
\rput[B](7,6.25){$b$}
\psline(3.2,3.2)(6,6)
\rput[B](7,-7.125){$c$}
\psline(-3.2,-3.2)(-6,-6)
\rput[B](-7,-7.125){$d$}
\psline(3.2,-3.2)(6,-6)
\psset{linecolor=gray,doubleline=false}
\psline(-3,3)(-3.2,3.2)
\psline(-3.2,3.2)(-6,6)
\psline(3,3)(3.2,3.2)
\psline(3.2,3.2)(6,6)
\psline(2.8,-2.8)(2.2,-2.2)
\psline(-3.2,-3.2)(-6,-6)
\psline(-2.8,-2.8)(-2.2,-2.2)
\psline(3.2,-3.2)(6,-6)
\end{pspicture}}
\subfigure[]{\label{fbox3}%
\begin{pspicture}(-8.5,-8.5)(8.5,8.5)
\recthreevertex{-3}{3}{135}
\threevertex{3}{3}{135}
\recthreevertex{3}{-3}{-45}
\threevertex{-3}{-3}{-45}
\psset{linecolor=black,doubleline=true}
\psbezier(-1.85,3)(0,3)(0,-3)(1.85,-3)
\psbezier(3,-1.85)(3,-0.85)(2.2929,1.7071)(1.2929,2.7071)
\psbezier(1.2929,2.7071)(0.2929,3.7071)(1.2929,4.7071)(2.2929,3.7071)
\psbezier(-2.2929,-3.7071)(-1.2929,-4.7071)(-0.2929,-3.7071)(-1.2929,-2.7071)
\psbezier(-1.2929,-2.7071)(-2.2929,-1.7071)(-3,0.85)(-3,1.85)

\rput[B](-7,6.25){$a$}
\psline(-3.2,3.2)(-6,6)
\rput[B](7,6.25){$b$}
\psline(3.7071,3.7071)(6,6)
\rput[B](7,-7.125){$c$}
\psline(3.2,-3.2)(6,-6)
\rput[B](-7,-7.125){$d$}
\psline(-3.7071,-3.7071)(-6,-6)
\psset{linecolor=gray,doubleline=false}
\psline(-3,3)(-3.2,3.2)
\psline(-3.2,3.2)(-6,6)
\psline(3,3)(3.7071,3.7071)
\psline(3.7071,3.7071)(6,6)
\psline(3,-3)(3.2,-3.2)
\psline(3.2,-3.2)(6,-6)
\psline(-3,-3)(-3.7071,-3.7071)
\psline(-3.7071,-3.7071)(-6,-6)
\psset{linecolor=lightgray,doubleline=false,linestyle=dashed}
\psline(-3,1.85)(-3,3)(-1.85,3)
\psbezier(-1.85,3)(0,3)(0,-3)(1.85,-3)
\psline(1.85,-3)(3,-3)(3,-1.85)
\psbezier(3,-1.85)(3,-0.85)(2.2929,1.7071)(1.2929,2.7071)
\psbezier(1.2929,2.7071)(0.2929,3.7071)(1.2929,4.7071)(2.2929,3.7071)
\psline(2.2929,3.7071)(3,3)(2.2929,2.2929)
\psline(-2.2929,-2.2929)(-3,-3)(-2.2929,-3.7071)
\psbezier(-2.2929,-3.7071)(-1.2929,-4.7071)(-0.2929,-3.7071)(-1.2929,-2.7071)
\psbezier(-1.2929,-2.7071)(-2.2929,-1.7071)(-3,0.85)(-3,1.85)

\psset{linecolor=black,doubleline=true,linestyle=solid}
\psline(-2.2929,-2.2929)(2.2929,2.2929)
\psset{linecolor=lightgray,doubleline=false,linestyle=dashed}
\psline(-2.2929,-2.2929)(2.2929,2.2929)
\end{pspicture}}
\caption{Complete list of contributions (up to conjugation) to 
$\bar\phi_i^a\bar\phi_i^b\phi^{i,c}\phi^{i,d}\big(ab\big)\big(cd\big)$ that deviate
from the ones in the undeformed $\mathcal{N}=4$ SYM theory.
The diagrams are displayed in double-line notation with central plain 
and dashed flavor lines for scalar and fermionic fields, respectively. 
Flavor-neutral gauge boson lines appear without central line.
\subref{ss4s4}, \subref{us4s4}: diagrams with two F-term-type single-trace interactions; 
\subref{ts4s4}: diagram with two F-term-type double-trace interactions; 
\subref{tDs4s4}: diagram with one F-term-type double-trace and one D-term-type single trace interaction; 
\subref{s4gg2}, \subref{s4gg3}:  F-term-type double-trace interaction with gauge boson exchange; 
\subref{fbox1}, \subref{fbox2}, \subref{fbox3}: fermion box with four 
Yukawa-type interactions. 
\label{fig:dtd}}
\end{center}
\end{figure}
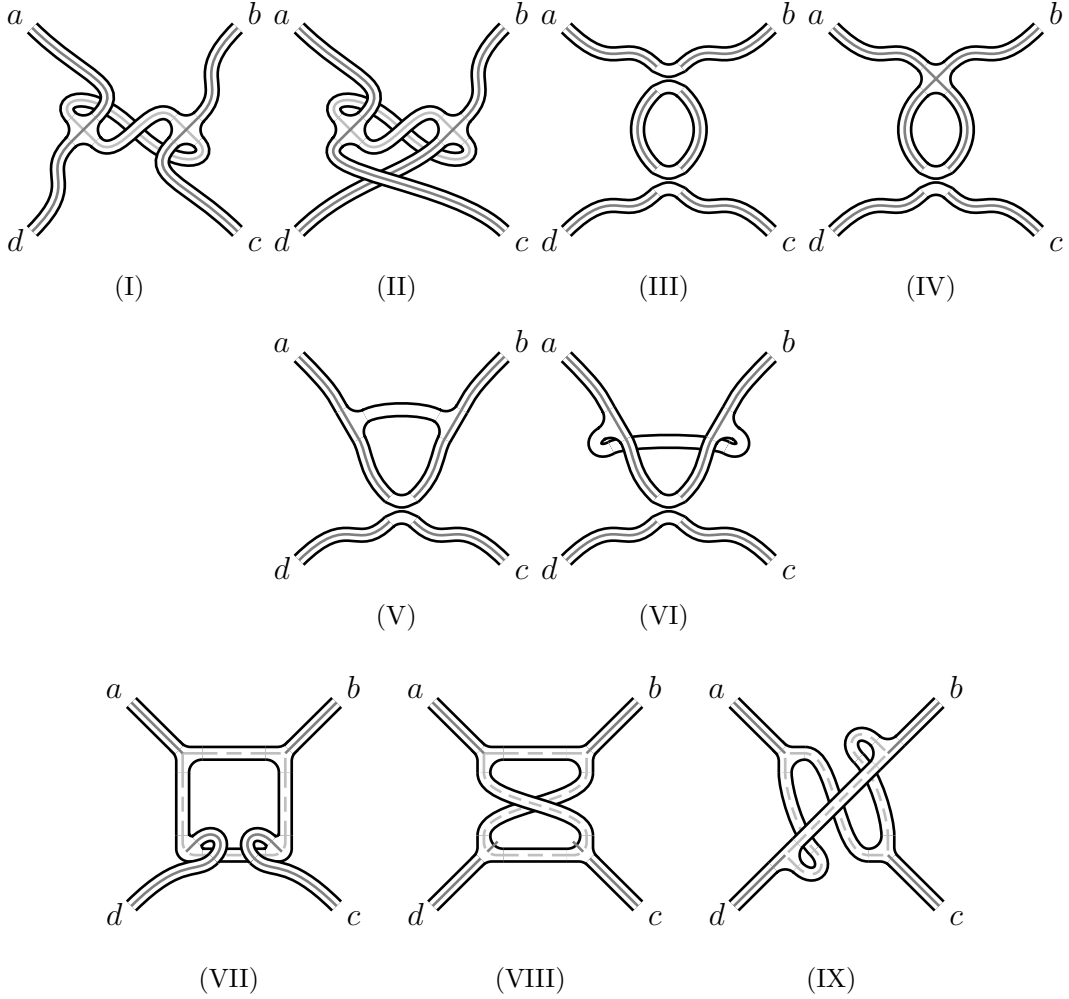

As mentioned in the previous section, in case of an $SU(N)$ gauge group, the only interactions that may supplement the $\gamma_i$-deformed action are the quartic double-trace terms \eqref{dtc} with two scalar fields in each trace. We use that in the $\mathcal{N}=4$ SYM theory all divergent contributions to the double-trace couplings vanish. This allows us to consider only those diagrams that are sensitive to the deformation and hence deviate from their $\mathcal{N}=4$ SYM theory counterparts. The deformation-dependent terms in the action of the $\gamma_i$-deformation \eqref{S} are the cubic Yukawa-type fermion-scalar couplings \eqref{rhodef} and the quartic F-term-type couplings of chiral and anti-chiral scalars \eqref{Fdef}. The only one-loop diagrams that depend on these couplings and that contribute at leading $N$-power to the interaction \eqref{zombiekiller} are displayed in Figure \ref{fig:dtd}.
Using the Feynman rules of Appendix \ref{app:frules} with unspecified 
gauge-fixing parameter $\alpha$,
these diagrams evaluate to
\begin{equation}
\begin{aligned}\label{diagres}
\subref{ss4s4}=\subref{us4s4}
&=g_\YM^4I_1\sum_{r=1}^3F^{ir}_{ri}F^{ir}_{ri}\big(ab\big)\big(cd\big)
\col\\
\Rop_|[\subref{ss4s4}]
=\Rop_|[\subref{us4s4}]
&=g_\YM^4I_1\sum_{r=1}^3F^{ri}_{ir}F^{ri}_{ir}\big(ab\big)\big(cd\big)
\col\\
\subref{ts4s4}
&=4g_\YM^4I_1\sum_{r,s=1}^3Q^{ii}_{\text{F}\,rs}(Q^{sr}_{\text{F}\,ii}+Q^{rs}_{\text{F}\,ii})\big(ab\big)\big(cd\big)
\col\\
\subref{tDs4s4}
&=4g_\YM^4I_1Q^{ii}_{\text{F}\,ii}\big(ab\big)\big(cd\big)
\col\\
\subref{s4gg2}=\subref{s4gg3}
&=2\alpha g_\YM^4I_1Q^{ii}_{\text{F}\,ii}\big(ab\big)\big(cd\big)
\col\\
\subref{fbox1}=\subref{fbox2}
&=-2g_\YM^4I_1
\big[\tr\big((\rho^{\dagger i})^{\T}(\tilde\rho^{\dagger i})^{\T}\tilde\rho_i\rho_i\big)
+\tr\big((\tilde\rho^{\dagger i})^{\T}(\rho^{\dagger i})^{\T}\rho_i\tilde\rho_i\big)\big]\big(ab\big)\big(cd\big)
\col\\
\subref{fbox3}&=
-2g_\YM^4I_1
\big[
\tr\big((\rho^{\dagger i})^{\T}\rho_i(\rho^{\dagger i})^{\T}\rho_i\big)
+\tr\big((\tilde\rho^{\dagger i})^{\T}\tilde\rho_i(\tilde\rho^{\dagger i})^{\T}\tilde\rho_i\big)\big]\big(ab\big)\big(cd\big)
\col\\
\end{aligned}
\end{equation}
where $F^{ij}_{lk}$ and $\rho_i$, $\tilde\rho_i$ are tensors 
of the quartic scalar F-term-type and cubic Yukawa couplings 
of the $\gamma_i$-deformed action \eqref{S}, respectively.
The operator $\Rop_|$ acts on a diagram by reflecting it at the vertical 
axis and restoring the original ordering of the labels at its external 
legs.
Similarly, some diagrams occur with
factors of two since an identical result coming from the diagram 
reflected at the horizontal axis has to be considered.
All contributions depend on a single scalar one-loop integral $I_1$ 
that is given by
\begin{equation}
I_1=\int\frac{\de^Dl}{(2\pi)^D}\frac{1}{l^2(p-l)^2}\col\qquad
\Kop[I_1]=\frac{1}{(4\pi)^2\varepsilon}
\pnt
\end{equation}
In the second equality, we have extracted the UV divergence of the integral by applying an operator $\Kop$. In dimensional reduction in $D=4-2\varepsilon$ dimensions, the UV divergences appear as poles in $\varepsilon$.

The individual sums of the UV divergences of all diagrams with only scalar interactions, with scalar and gauge-boson interactions and with a fermion loop are given by
\begin{equation}
\begin{aligned}\label{dtdsubsum}
\Kop\left[(1+\Rop_|)[\,\subref{ss4s4}+\subref{us4s4}\,]+\subref{ts4s4}+2\,\subref{tDs4s4}\right]
&=8g_\YM^4\Kop[I_1]\big(
\cos^2\epsilon_{ijk}\gamma_j+\cos^2\epsilon_{ijk}\gamma_k-1\\
&\hphantom{{}={}8g_\YM^4\Kop[I_1]\big(}
+(Q^{ii}_{\text{F}\,ii})^2+Q^{ii}_{\text{F}\,ii}\big)\big(ab\big)\big(cd\big)
\col\\
2\Kop\left[\subref{s4gg2}+\subref{s4gg3}\right]
&=8\alpha g_\YM^4\Kop[I_1]Q^{ii}_{\text{F}\,ii}\big(ab\big)\big(cd\big)
\col\\
\Kop\left[(1+\Rop_|)\subref{fbox3}\right]
&=-16g_\YM^4\Kop[I_1]\cos\epsilon_{ijk}\gamma_j\cos\epsilon_{ijk}\gamma_k
\big(ab\big)\big(cd\big)
\col
\end{aligned}
\end{equation}
where in the first line we have used \eqref{FFsum} and the conservation of the global $U(1)$ charges for the coupling $Q^{ij}_{\text{F}\,lk}$, and in the last line we have inserted \eqref{l4rhotraces}. Note that in the above expressions we do \emph{not} make use of Einstein's summation convention. Instead, the resulting expressions that contain $\epsilon_{ijk}$ have to be evaluated fixing $i$, $j$, $k$ to one of the three cyclic permutations $(i,j,k)\in\{(1,2,3),(2,3,1),(3,1,2)\}$. We still have to reconstruct the divergent contributions to the double-trace coupling \eqref{zombiekiller} that come from the neglected deformation-independent diagrams. To this purpose we use the aforementioned fact that in the $\mathcal{N}=4$ SYM theory the sum of all divergent contributions to \eqref{zombiekiller} has to vanish. Sending the deformation parameters $\gamma_i$ and the tree-level couplings $Q^{ii}_{\text{F}\,ii}$ to zero, the deformation-independent diagrams are not altered and the deformation-dependent contributions in \eqref{dtdsubsum} reduce to the respective $\mathcal{N}=4$ SYM results. This yields $8g_\YM^4\Kop[I_1]$ for the diagrams with two quartic scalar vertices and $-16g_\YM^4\Kop[I_1]$ for the ones with a fermion-loop. The UV divergence from all diagrams in the $\gamma_i$-deformation is obtained by subtracting these results from the sum of all expressions in \eqref{dtdsubsum}. It yields the following counter term for the coupling \eqref{zombiekiller}:
\begin{equation}
\begin{aligned}\label{deltaQ}
\delta Q^{ii}_{\text{F}\,ii}
&=-\Kop\left.\left[
\qvertr[
\fmfiv{label=$\scriptstyle i a$,l.dist=2}{vloc(__v1)}
\fmfiv{label=$\scriptstyle i b$,l.dist=2}{vloc(__v2)}
\fmfiv{label=$\scriptstyle i c$,l.dist=2}{vloc(__v3)}
\fmfiv{label=$\scriptstyle i d$,l.dist=2}{vloc(__v4)}
\fmfcmd{fill(fullcircle scaled 10 shifted vloc(__vc1)) withcolor 0.2black;}
]{plain_rar,l.side=left,l.dist=2,l.dist=2}{plain_rar,l.side=left,l.dist=2,l.dist=2}{plain_ar,l.side=left,l.dist=2,l.dist=2}{plain_ar,l.side=left,l.dist=2,l.dist=2}{}{}{}{}
\right]\right|_{-4\frac{g_\YM^2}{N}(ab)(cd)}
\\
&=2\frac{g_\YM^2N}{(4\pi)^2\varepsilon}\big((\cos\epsilon_{ijk}\gamma_j-\cos\epsilon_{ijk}\gamma_k)^2
+(Q^{ii}_{\text{F}\,ii})^2-(1+\alpha)Q^{ii}_{\text{F}\,ii}\big)
\col
\end{aligned}
\end{equation}
where the vertical bar indicates that the coefficient of the specified expression -- the term multiplying $\delta Q^{ii}_{\text{F}\,ii}$ in the Feynman rule for the counter term -- is taken. Together with the respective tree-level coupling the counter term enters the action as
\begin{equation}
\label{zombiekillerct}
-\frac{g_\YM^2}{N}(Q^{ii}_{\text{F}\,ii}+\delta Q^{ii}_{\text{F}\,ii})\tr(\bar\phi_i\bar\phi_i)\tr(\phi^i\phi^i)
\pnt
\end{equation}

In order to obtain the renormalization of the corresponding coupling, we have to add contributions from wave function renormalization, as reviewed in Appendix \ref{app:couplren}. They come from diagrams that involve a tree-level quartic scalar vertex with a self-energy correction at one of its external legs. In these diagrams, the only sources for double-trace terms are the quartic scalar double-trace couplings themselves. This follows from the fact that a self-energy correction at one of the external legs of a quartic vertex cannot generate traces of two fields by itself. Connecting the divergent diagrams of the self-energy corrections \eqref{sephi} to the vertex \eqref{zombiekiller}, the relevant diagrams contribute as
\begin{equation}
\begin{aligned}\label{selfenergycontrib}
&-\frac 12\Kop\left.\left[
\qvertr[
\fmfiv{label=$\scriptstyle i a$,l.dist=2}{vloc(__v1)}
\fmfiv{label=$\scriptstyle i b$,l.dist=2}{vloc(__v2)}
\fmfiv{label=$\scriptstyle i c$,l.dist=2}{vloc(__v3)}
\fmfiv{label=$\scriptstyle i d$,l.dist=2}{vloc(__v4)}
\fmfcmd{pair vert[]; vert1 = vloc(__vc1);}
\fmfi{plain_ar}{vert1--vi1}
\fmfi{plain_ar}{vo1--vloc(__v1)}
\fmfcmd{fill(fullcircle scaled 7.5 shifted vm1) withcolor 0.2black;}
]{phantom}{plain_rar}{plain_ar}{plain_ar}{}{}{}{}
+
\qvertr[
\fmfiv{label=$\scriptstyle i a$,l.dist=2}{vloc(__v1)}
\fmfiv{label=$\scriptstyle i b$,l.dist=2}{vloc(__v2)}
\fmfiv{label=$\scriptstyle i c$,l.dist=2}{vloc(__v3)}
\fmfiv{label=$\scriptstyle i d$,l.dist=2}{vloc(__v4)}
\fmfcmd{pair vert[]; vert1 = vloc(__vc1);}
\fmfi{plain_ar}{vert1--vi2}
\fmfi{plain_ar}{vo2--vloc(__v2)}
\fmfcmd{fill(fullcircle scaled 7.5 shifted vm2) withcolor 0.2black;}
]{plain_rar}{phantom}{plain_ar}{plain_ar}{}{}{}{}
+
\qvertr[
\fmfiv{label=$\scriptstyle i a$,l.dist=2}{vloc(__v1)}
\fmfiv{label=$\scriptstyle i b$,l.dist=2}{vloc(__v2)}
\fmfiv{label=$\scriptstyle i c$,l.dist=2}{vloc(__v3)}
\fmfiv{label=$\scriptstyle i d$,l.dist=2}{vloc(__v4)}
\fmfcmd{pair vert[]; vert1 = vloc(__vc1);}
\fmfi{plain_rar}{vert1--vi3}
\fmfi{plain_rar}{vo3--vloc(__v3)}
\fmfcmd{fill(fullcircle scaled 7.5 shifted vm3) withcolor 0.2black;}
]{plain_rar}{plain_rar}{phantom}{plain_ar}{}{}{}{}
+
\qvertr[
\fmfiv{label=$\scriptstyle i a$,l.dist=2}{vloc(__v1)}
\fmfiv{label=$\scriptstyle i b$,l.dist=2}{vloc(__v2)}
\fmfiv{label=$\scriptstyle i c$,l.dist=2}{vloc(__v3)}
\fmfiv{label=$\scriptstyle i d$,l.dist=2}{vloc(__v4)}
\fmfcmd{pair vert[]; vert1 = vloc(__vc1);}
\fmfi{plain_rar}{vert1--vi4}
\fmfi{plain_rar}{vo4--vloc(__v4)}
\fmfcmd{fill(fullcircle scaled 7.5 shifted vm4) withcolor 0.2black;}
]{plain_rar}{plain_rar}{plain_ar}{phantom}{}{}{}{}\right]\right|_{-4\frac{g_\YM^2}{N}(ab)(cd)}\\
&=
-2\Kop\Big[
\settoheight{\eqoff}{$\times$}%
\setlength{\eqoff}{0.5\eqoff}%
\addtolength{\eqoff}{-7.5\unitlength}%
\raisebox{\eqoff}{%
\fmfframe(1,0)(1,0){%
\begin{fmfchar*}(20,15)
\fmfleft{v1}
\fmfright{v2}
\fmf{plain_ar}{v1,vl}
\fmf{phantom}{vl,vr}
\fmf{plain_ar}{vr,v2}
\fmffreeze
\fmfposition
\fmfcmd{pair vert[]; vert1 = vloc(__vl);}
\vacpolp[1]{vert1--vloc(__vr)}
\end{fmfchar*}}}
\Big]\frac{1}{p^2}\!\!\!
\left.
\qvertr[
\fmfiv{label=$\scriptstyle i a$,l.dist=2}{vloc(__v1)}
\fmfiv{label=$\scriptstyle i b$,l.dist=2}{vloc(__v2)}
\fmfiv{label=$\scriptstyle i c$,l.dist=2}{vloc(__v3)}
\fmfiv{label=$\scriptstyle i d$,l.dist=2}{vloc(__v4)}
]{plain_rar}{plain_rar}{plain_ar}{plain_ar}{}{}{}{}\right|_{-4\frac{g_\YM^2}{N}(ab)(cd)}
\!\!=-2\delta_{\phi^i} Q^{ii}_{\text{F}\,ii}
=2(1+\alpha)\frac{g_\YM^2N}{(4\pi)^2\varepsilon}Q^{ii}_{\text{F}\,ii}
\col
\end{aligned}
\end{equation}
where $\delta_{\phi^i}$ is the wave function renormalization counter term for the $SU(N)$ components of the scalar fields, explicitly given in \eqref{deltaphi}. The coupling renormalization is then given by the sum of \eqref{deltaQ} and \eqref{selfenergycontrib}
\begin{equation}
\begin{aligned}
\label{1111dtcoupren}
Q^{ii}_{\text{F}\,ii}\delta_{Q_{\text{F}\,ii}^{ii}}&=\delta Q^{ii}_{\text{F}\,ii}-2\delta_{\phi^i} Q^{ii}_{\text{F}\,ii}
=2\frac{g_\YM^2N}{(4\pi)^2\varepsilon}\big((\cos\epsilon_{ijk}\gamma_j-\cos\epsilon_{ijk}\gamma_k)^2
+(Q^{ii}_{\text{F}\,ii})^2\big)
\col
\end{aligned}
\end{equation}
where the dependence on the gauge-fixing parameter $\alpha$ has canceled as required.\footnote{Note that \eqref{1111dtcoupren} does not contain a contribution linear in the double-trace coupling: the respective gauge-dependent terms in the coupling counter term are canceled by the diagrams containing the self-energy correction. This is in accord with the observations in \cite{Dymarsky:2005uh} for orbifolds: the term linear in the double-trace coupling is proportional to the anomalous dimension of its single-trace factor. Since the one-loop anomalous dimension of $\tr(\phi^i\phi^i)$ is proportional to $Q^{ii}_{\text{F}\,ii}$ itself \cite{Fokken:2014soa}, \eqref{1111dtcoupren} does not contain a term linear in $Q^{ii}_{\text{F}\,ii}$.} The above result agrees with unpublished results of \cite{DymarskyRoiban}.\footnote{We thank Radu Roiban for communication on this point.} Since the conditions \eqref{conjQFD} hold, the second term in parenthesis is positive, as is the first one, and the coupling is not renormalized only if all $\gamma_i$, $i=1,2,3$ are identical (modulo shifts by integer multiples of $\pi$ and signs factors). Hence, for generic angles $\gamma_i$ subject to the conditions formulated in Subsection \ref{subsec:conclusions}, the one-loop coupling renormalization leads to a non-vanishing $\beta$-function, and conformal invariance is broken. Using the expression \eqref{betaQlo}, the $\beta$-function for the coupling $Q^{ii}_{\text{F}\,ii}$ reads
\begin{equation}
\begin{aligned}\label{betaQres}
\beta_{Q_{\text{F}\,ii}^{ii}}
=\varepsilon g_\YM\parderiv{g_\YM}Q_{\text{F}\,ii}^{ii}\delta_{Q_{\text{F}\,ii}^{ii}}
=4\frac{g_\YM^2N}{(4\pi)^2}\big((\cos\epsilon_{ijk}\gamma_j-\cos\epsilon_{ijk}\gamma_k)^2
+(Q^{ii}_{\text{F}\,ii})^2\big)
\pnt
\end{aligned}
\end{equation}

\subsection{\texorpdfstring{$U(N)$ gauge group}{U(N) gauge group}}

In case of the $U(N)$ gauge group, the additional couplings \eqref{cUNc}, \eqref{qUNc} could in principle alter \eqref{1111dtcoupren} and hence \eqref{betaQres} such that a non-running double-trace coupling for the $SU(N)$ components is possible. 

The additional Feynman diagrams are given by replacing the vertices in Figure \ref{fig:dtd} by the respective ones obtained from \eqref{cUNc} and \eqref{qUNc}, while keeping  the double-trace structure of the external lines intact. The reader may convince herself that all these diagrams are suppressed by powers of $\frac{1}{N}$, since the vertices with enhanced numbers of traces cannot increase the number of internal color loops but come with additional factors of $\frac{1}{N}$.

Thus, in the 't Hooft limit, the results in Subsection \ref{subsec:SUN} for the coupling \eqref{zombiekiller} are not affected by the additional couplings \eqref{cUNc} and \eqref{qUNc}; they remain valid for the $SU(N)$ components in the $U(N)$ theory.

\section*{Acknowledgements}

We are very grateful to Sergey Frolov and Radu Roiban for enlightening discussions, reading the manuscript and giving feedback for improvements. Moreover, we thank Zoltan Bajnok, Stefano Kovacs and Stijn van Tongeren for intense discussions. We also thank Matthias Staudacher for discussions in the initial phase of this project and for comments on the manuscript. Our work was supported by DFG, SFB 647 \emph{Raum -- Zeit -- Materie. Analytische und Geometrische Strukturen}. J.F.\ and M.W.\ danken der Studienstiftung des deutschen Volkes f\"ur ein Promotionsf\"orderungsstipendium.

\appendix
\section{\texorpdfstring{The action of $\gamma_i$-deformed $\mathcal{N}=4$ SYM 
theory}{The action of gamma\_i-deformed N=4 SYM theory}}
\label{app:action}

In this appendix, we present the $\gamma_i$-deformation as well as our notation and conventions. The gauge-fixed action of ($\gamma_i$-deformed) $\mathcal{N}=4$ SYM theory in Euclidean space can be written as
\footnote{Note that we have included a double-trace coupling which for $\gamma_1=\gamma_2=\gamma_3=-\pi\beta$ reduces to the one already present in the conformal $\beta$-deformation with gauge group $SU(N)$, cf.\ Section \ref{sec:doubletrcoupl}. In addition, the action has to be supplemented with the multi-trace couplings of Section \ref{sec:multitrdef}.}
\begin{equation}
\begin{aligned}\label{S}
S
&=\int\de^4x\,\Big[\tr\Big(
-\frac{1}{4}F^{\mu\nu}F_{\mu\nu}-\frac{1}{2\alpha}(\partial^\mu A_\mu)^2
-(\D^\mu\bar\phi_i)\D_\mu\phi^i
+i\,\bar\psi^{\dot\alpha}_A(\tilde\sigma^\mu)_{\dot\alpha}{}^\alpha\D_\mu\psi_\alpha^A\\
&\hphantom{{}={}\int\de^4x\,\Big[\tr\Big(}
+g_\YM((\tilde\rho_i)^{BA}\bar\psi^{\dot\alpha}_A\phi^i\bar\psi_{\dot\alpha\,B}
+(\tilde\rho^{\dagger\,i})_{BA}\psi^{\alpha\,A}\bar\phi_i\psi_\alpha^B)\\
&\phantom{{}={}\int\de^4x\,\Big[\tr\Big(}
+g_\YM((\rho_{i})_{BA}\psi^{\alpha\,A}\phi^i\psi_\alpha^B
+(\rho^{\dagger\,i})^{BA}\bar\psi^{\dot\alpha}_A\bar\phi_i\bar\psi_{\dot\alpha\,B})
+\bar c\,\partial^\mu\D_\mu c\\
&\phantom{{}={}\int\de^4x\,\Big[\tr\Big(}
-\frac{g_\YM^2}{4}\comm{\bar\phi_i}{\phi^i}\comm{\bar\phi_j}{\phi^j}
+g_\YM^2F_{lk}^{ij}\,\bar\phi_i\bar\phi_j\phi^k\phi^l\Big)\\
&\phantom{{}={}\int\de^4x\,\Big[}
-g_\YM^2\frac{s}{N}F_{lk}^{ij}\tr(\bar\phi_i\bar\phi_j)\tr(\phi^k\phi^l)\Big]\col
\end{aligned}
\end{equation}
where we have adopted the conventions of \cite{Gates:1983nr}, in particular the ones for raising, lowering and the contractions of the spinor indices $\alpha,\dot\alpha=1,2$. Moreover, we have $(\tilde\sigma_\mu)_{\dot\alpha}{}^\alpha=(-i\sigma_2,i\sigma_3,\unitmatrix,-i\sigma_1)_{\dot\alpha}{}^\alpha$ in terms of the identity $\unitmatrix$ and the Pauli matrices $\sigma_k$. The covariant derivative acts on the chiral and anti-chiral scalar fields  $\phi^i$ and $\bar\phi_i$ ($i=1,2,3$), vectors $A_\mu$, ghosts $c$ and spinors $\psi_\alpha^A$ ($A=1,2,3,4$), as
\begin{equation}
\begin{aligned}
\D_\mu&=\partial_\mu+i\frac{g_\YM}{\sqrt{2}}\comm{A_\mu}{\cdot}\pnt
\end{aligned}
\end{equation}
It determines the Yang-Mills field strength as follows: 
\begin{equation}
\begin{aligned}
F_{\mu\nu}&=-i\frac{\sqrt{2}}{g_\YM}\comm{\D_\mu}{\D_\nu}
=\partial_\mu A_\nu-\partial_\nu A_\mu+i\frac{g_\YM}{\sqrt{2}}\comm{A_\mu}{A_\nu}
\pnt
\end{aligned}
\end{equation}
All fields in the action \eqref{S} transform in the adjoint representation of the $SU(N)$ or $U(N)$ gauge group. The representation matrices obey the relations \eqref{TTprod}. 

The Yukawa and quartic scalar F-term-type couplings in the action \eqref{S} are subject to the $\gamma_i$-deformation, which introduces phase factors depending on the three deformation angles $\gamma_i$, $i=1,2,3$, into the couplings.

As mentioned in Section \ref{sec:doubletrcoupl}, the deformed action \eqref{S} is obtained from the $\mathcal{N}=4$ SYM action in component fields by replacing all products of fields by $\ast$-products before integrating out the auxiliary fields. The $\ast$-product of two component fields $A$ and $B$ reads
\begin{equation}\label{astproddef}
A\ast B=\e^{\frac{i}{2}\mathbf{q}_A\wedge\mathbf{q}_B} AB\col
\end{equation}
where the antisymmetric product of the two charge vectors $\mathbf{q}_A$ and $\mathbf{q}_B$ is given by
\begin{equation}
\mathbf{q}_A\wedge \mathbf{q}_B=(\mathbf{q}_A)^{\Top}\mathbf{C}\,\mathbf{q}_B
\col\qquad
\mathbf{C}=\begin{pmatrix}
0 & -\gamma_3 & \gamma_2 \\
\gamma_3 & 0 & -\gamma_1 \\
-\gamma_2 & \gamma_1 & 0 
\end{pmatrix}
\pnt
\end{equation} 
The $U(1)\times U(1)\times U(1)$ charge vectors $\mathbf{q}_B=(q_B^1,q_B^2,q_B^3)^{\Top}$ of the component fields are given by 
\begin{equation}
\begin{array}{c|cccc|c|ccc}
B&\psi^1_{\alpha}&\psi^2_{\alpha}&\psi^3_{\alpha}&\psi^4_{\alpha}
&A_{\mu}&\phi_1&\phi_2&\phi_3\\
 & & & & & & & &\\[-0.4cm]
\hline
 & & & & & & & &\\[-0.4cm]
q^1_B & +\frac12 & -\frac12 & -\frac12 & +\frac12 & 0 & 1 & 0 & 0\\
 & & & & & & & &\\[-0.4cm]
q^2_B & -\frac12 & +\frac12 & -\frac12 & +\frac12 & 0 & 0 & 1 & 0\\
 & & & & & & & &\\[-0.4cm]
q^3_B & -\frac12 & -\frac12 & +\frac12 & +\frac12 & 0 & 0 & 0 & 1\\
\end{array}
\pnt
\end{equation}
Respective relations hold for the anti-fields with reversed charge vectors. We define the following abbreviations for the independent components:
\begin{equation}
\begin{aligned}
\Gamma_{i4}
&=\mathbf{q}_{\psi^i}\wedge\mathbf{q}_{\psi^4}
=\frac{1}{4}\sum_{j,k=1}^3\epsilon_{ijk}(\gamma_j-\gamma_k)
=\frac{1}{2}\epsilon_{ijk}(\gamma_j-\gamma_k)
\col\\
\Gamma_{ij}
&=\mathbf{q}_{\psi^i}\wedge\mathbf{q}_{\psi^j}
=
-\frac{1}{2}\sum_{k=1}^3\epsilon_{ijk}(\gamma_i+\gamma_j)
=-\frac{1}{2}\epsilon_{ijk}(\gamma_i+\gamma_j)\col\\
\Gamma^+_{ij}
&=\mathbf{q}_{\phi^i}\wedge\mathbf{q}_{\phi^j}
=-\epsilon_{ijk}\gamma_k
\col
\end{aligned}
\end{equation}
where we interpret the expressions without Einstein's summation convention. Instead, the index $i=1,2,3$ is fixed and $j$ and $k$ assume the values of the corresponding cyclic permutation $(i,j,k)\in\{(1,2,3),(2,3,1),(3,1,2)\}$ in the results.

In terms of the fermionic phase tensor $\Gamma_{AB}$, the Yukawa coupling tensors in the action \eqref{S} are explicitly given by 
\begin{equation}\label{rhodef}
(\rho_{i})_{AB}=i\epsilon_{4iAB}\e^{\frac{i}{2}\Gamma_{AB}}
\col\qquad
(\tilde\rho_i)^{AB}=(\delta_4^A\delta_i^B-\delta_4^B\delta_i^A)\e^{\frac{i}{2}\Gamma_{AB}}
\col
\end{equation}
and they obey the conjugation relations
\begin{equation}
\begin{aligned}
(\rho^{\dagger i})^{AB}
&=((\rho_{i})_{BA})^\ast
=(\rho_{i})_{AB}
\col\qquad
(\tilde\rho^{\dagger i})_{AB}
=
((\tilde\rho_i)^{BA})^\ast
=-(\tilde\rho_i)^{AB}\pnt
\end{aligned}
\end{equation}

Moreover, the deformation enters the F-term-type coupling tensor 
via the bosonic phase tensor $\Gamma^+_{ij}$
as follows:
\begin{equation}
\begin{aligned}\label{Fdef}
F_{lk}^{ij}
&=
\delta_k^i\delta_l^j-\delta_l^i\delta_k^j
+Q_{lk}^{ij}
\col\qquad
Q_{lk}^{ij}=\delta_k^i\delta_l^j(\e^{i\Gamma^+_{ij}}-1)\col
\end{aligned}
\end{equation}
where we have split the coupling tensor into the F-term tensor of the 
undeformed $\mathcal{N}=4$ SYM theory and 
a tensor $Q_{lk}^{ij}$ carrying the deformation.
Reality of the action requires that the tensor $F_{lk}^{ij}$ 
(and hence also $Q_{lk}^{ij}$) obeys the conjugation relation
\begin{equation}\label{Fbarrel}
(F_{lk}^{ij})^\ast=(F_{ij}^{lk})\pnt
\end{equation}

\section{Feynman rules}
\label{app:frules}
In this appendix, we list the Feynman rules of the $\gamma_i$-deformation.
The propagators are given as the negative of the inverse 
kernels as extracted from the terms in \eqref{S} that are quadratic in the fields.
In our conventions, a transformation to momentum space is simply performed by 
replacing $i\partial_\mu\to p_\mu$ when $p_\mu$ leaves the 
vertex. One obtains the expressions
\begin{equation}\label{propagators}
\begin{aligned}
\settoheight{\eqoff}{$\times$}%
\setlength{\eqoff}{0.5\eqoff}%
\addtolength{\eqoff}{-5.75\unitlength}%
\raisebox{\eqoff}{%
\fmfframe(5,2)(5,2){%
\begin{fmfchar*}(15,7.5)
\fmfleft{v1}
\fmfright{v2}
\fmfforce{0.0625w,0.5h}{v1}
\fmfforce{0.9375w,0.5h}{v2}
\fmf{photon}{v1,v2}
\fmffreeze
\fmfposition
\fmfipath{pm[]}
\fmfiset{pm1}{vpath(__v1,__v2)}
\nvml{1}{$\scriptstyle p$}
\fmfiv{label=$\scriptstyle \nu b$,l.dist=2}{vloc(__v1)}
\fmfiv{label=$\scriptstyle \mu a$,l.dist=2}{vloc(__v2)}
\end{fmfchar*}}}
&{}={}
\langle A^{\mu a}(-p)A^{\nu b}(p)\rangle
=\frac{1}{p^2}\Big(g^{\mu\nu}-(1-\alpha)\frac{p^\mu p^\nu}{p^2}\Big)\delta^{ab}
\col\\
\settoheight{\eqoff}{$\times$}%
\setlength{\eqoff}{0.5\eqoff}%
\addtolength{\eqoff}{-5.75\unitlength}%
\raisebox{\eqoff}{%
\fmfframe(5,2)(5,2){%
\begin{fmfchar*}(15,7.5)
\fmfleft{v1}
\fmfright{v2}
\fmfforce{0.0625w,0.5h}{v1}
\fmfforce{0.9375w,0.5h}{v2}
\fmf{plain_ar}{v1,v2}
\fmffreeze
\fmfposition
\fmfipath{pm[]}
\fmfiset{pm1}{vpath(__v1,__v2)}
\nvml{1}{$\scriptstyle p$}
\fmfiv{label=$\scriptstyle jb$,l.dist=2}{vloc(__v1)}
\fmfiv{label=$\scriptstyle ia$,l.dist=2}{vloc(__v2)}
\end{fmfchar*}}}
&{}={}
\langle \phi^{ia}(-p)\bar\phi_j^b(p)\rangle
=\frac{1}{p^2}\delta_j^i\delta^{ab}
\col\\
\settoheight{\eqoff}{$\times$}%
\setlength{\eqoff}{0.5\eqoff}%
\addtolength{\eqoff}{-5.75\unitlength}%
\raisebox{\eqoff}{%
\fmfframe(5,2)(5,2){%
\begin{fmfchar*}(15,7.5)
\fmfleft{v1}
\fmfright{v2}
\fmfforce{0.0625w,0.5h}{v1}
\fmfforce{0.9375w,0.5h}{v2}
\fmf{dashes_ar}{v1,v2}
\fmffreeze
\fmfposition
\fmfipath{pm[]}
\fmfiset{pm1}{vpath(__v1,__v2)}
\nvml{1}{$\scriptstyle p$}
\fmfiv{label=$\scriptstyle {\dot\beta}Bb$,l.dist=2}{vloc(__v1)}
\fmfiv{label=$\scriptstyle  \alpha Aa$,l.dist=2}{vloc(__v2)}
\end{fmfchar*}}}
&{}={}
\langle\psi^{Aa}_\alpha(-p)\bar\psi_B^{\dot\beta b}(p)\rangle
=-\delta_B^A\delta^{ab}\frac{p_\alpha{}^{\dot\beta}}{p^2}\col\\
\settoheight{\eqoff}{$\times$}%
\setlength{\eqoff}{0.5\eqoff}%
\addtolength{\eqoff}{-5.75\unitlength}%
\raisebox{\eqoff}{%
\fmfframe(5,2)(5,2){%
\begin{fmfchar*}(15,7.5)
\fmfleft{v1}
\fmfright{v2}
\fmfforce{0.0625w,0.5h}{v1}
\fmfforce{0.9375w,0.5h}{v2}
\fmf{dots}{v1,v2}
\fmffreeze
\fmfposition
\fmfipath{pm[]}
\fmfiset{pm1}{vpath(__v1,__v2)}
\nvml{1}{$\scriptstyle p$}
\fmfiv{label=$\scriptstyle a$,l.dist=2}{vloc(__v1)}
\fmfiv{label=$\scriptstyle b$,l.dist=2}{vloc(__v2)}
\end{fmfchar*}}}
&{}={}
\langle c^a(-p)\bar c^b(p)\rangle
=\frac{1}{p^2}\delta^{ab}\pnt
\end{aligned}
\end{equation}

The vertices are obtained from \eqref{S} by taking the functional derivatives  w.r.t.\ the corresponding fields. We obtain for the cubic vertices
\begin{equation}\label{cvertices}
\begin{gathered}
\begin{aligned}
V_{AAA}
&=
\cvertr[
\fmfiv{label=$\scriptstyle \mu a$,l.dist=2}{vloc(__v1)}
\fmfiv{label=$\scriptstyle \nu b$,l.dist=2}{vloc(__v2)}
\fmfiv{label=$\scriptstyle \rho c$,l.dist=2}{vloc(__v3)}
]{photon,label=$\scriptstyle p$,l.side=left,l.dist=2,l.dist=2}{photon,label=$\scriptstyle q$,l.side=right,l.dist=2,l.dist=2}{photon,label=$\scriptstyle r$,l.side=right,l.dist=2,l.dist=2}
=\frac{g_\YM}{\sqrt{2}}\big[(p-q)_\rho g_{\mu\nu}+(q-r)_\mu g_{\nu\rho}+(r-p)_\nu g_{\rho\mu}\big]\big(a\comm{b}{c}\big)
\col\\
V_{\bar\phi A\phi}
&=
\cvertr[
\fmfiv{label=$\scriptstyle ia$,l.dist=2}{vloc(__v1)}
\fmfiv{label=$\scriptstyle \nu b$,l.dist=2}{vloc(__v2)}
\fmfiv{label=$\scriptstyle kc$,l.dist=2}{vloc(__v3)}
]{plain_rar,label=$\scriptstyle p$,l.side=left,l.dist=2,l.dist=2}{photon,label=$\scriptstyle q$,l.side=right,l.dist=2,l.dist=2}{plain_ar,label=$\scriptstyle r$,l.side=right,l.dist=2,l.dist=2}
=-\frac{g_\YM}{\sqrt{2}}(p-r)_\nu\delta^i_k\big(a\comm{b}{c}\big)
\col\\
V_{\bar\psi A\psi}
&=
\cvertr[
\fmfiv{label=$\scriptstyle \dot\alpha Aa$,l.dist=2}{vloc(__v1)}
\fmfiv{label=$\scriptstyle \nu b$,l.dist=2}{vloc(__v2)}
\fmfiv{label=$\scriptstyle \gamma Cc$,l.dist=2}{vloc(__v3)}
]{dashes_rar,label=$\scriptstyle p$,l.side=left,l.dist=2,l.dist=2}{photon,label=$\scriptstyle q$,l.side=right,l.dist=2,l.dist=2}{dashes_ar,label=$\scriptstyle r$,l.side=right,l.dist=2,l.dist=2}
=-\frac{g_\YM}{\sqrt{2}}(\tilde\sigma_\nu)_{\dot\alpha}{}^\gamma
\delta^A_C\big(a\comm{b}{c}\big)
\col\\
V_{\psi\phi\psi}
&=
\cvertr[
\fmfiv{label=$\scriptstyle \alpha A a$,l.dist=2}{vloc(__v1)}
\fmfiv{label=$\scriptstyle jb$,l.dist=2}{vloc(__v2)}
\fmfiv{label=$\scriptstyle \gamma C c$,l.dist=2}{vloc(__v3)}
]{dashes_ar,label=$\scriptstyle p$,l.side=left,l.dist=2,l.dist=2}{plain_ar,label=$\scriptstyle q$,l.side=right,l.dist=2,l.dist=2}{dashes_ar,label=$\scriptstyle r$,l.side=right,l.dist=2,l.dist=2}
=g_\YM\delta_\alpha{}^\gamma\big[(\rho_{j})_{CA}\big(abc\big)+(\rho_{j})_{AC}\big(acb\big)\big]
\col\\
V_{\bar\psi\bar\phi\bar\psi}
&=
\cvertr[
\fmfiv{label=$\scriptstyle \dot\alpha A a$,l.dist=2}{vloc(__v1)}
\fmfiv{label=$\scriptstyle jb$,l.dist=2}{vloc(__v2)}
\fmfiv{label=$\scriptstyle \dot\gamma C c$,l.dist=2}{vloc(__v3)}
]{dashes_rar,label=$\scriptstyle p$,l.side=left,l.dist=2,l.dist=2}{plain_rar,label=$\scriptstyle q$,l.side=right,l.dist=2,l.dist=2}{dashes_rar,label=$\scriptstyle r$,l.side=right,l.dist=2,l.dist=2}
=g_\YM\delta_{\dot\alpha}{}^{\dot\gamma}\big[(\rho^{\dagger\,j})^{CA}\big(abc\big)
+(\rho^{\dagger\,j})^{AC}\big(acb\big)\big]
\col\\
V_{\psi\bar\phi\psi}
&=
\cvertr[
\fmfiv{label=$\scriptstyle \alpha A a$,l.dist=2}{vloc(__v1)}
\fmfiv{label=$\scriptstyle jb$,l.dist=2}{vloc(__v2)}
\fmfiv{label=$\scriptstyle \gamma C c$,l.dist=2}{vloc(__v3)}
]{dashes_ar,label=$\scriptstyle p$,l.side=left,l.dist=2,l.dist=2}{plain_rar,label=$\scriptstyle q$,l.side=right,l.dist=2,l.dist=2}{dashes_ar,label=$\scriptstyle r$,l.side=right,l.dist=2,l.dist=2}
=g_\YM\delta_\alpha{}^\gamma\big[(\tilde\rho^{\dagger\,j})_{CA}\big(abc\big)
+(\tilde\rho^{\dagger\,j})_{AC}\big(acb\big)\big]
\col\\
V_{\bar\psi\phi\bar\psi}
&=
\cvertr[
\fmfiv{label=$\scriptstyle \dot\alpha A a$,l.dist=2}{vloc(__v1)}
\fmfiv{label=$\scriptstyle jb$,l.dist=2}{vloc(__v2)}
\fmfiv{label=$\scriptstyle \dot\gamma C c$,l.dist=2}{vloc(__v3)}
]{dashes_rar,label=$\scriptstyle p$,l.side=left,l.dist=2,l.dist=2}{plain_ar,label=$\scriptstyle q$,l.side=right,l.dist=2,l.dist=2}{dashes_rar,label=$\scriptstyle r$,l.side=right,l.dist=2,l.dist=2}
=g_\YM\delta_{\dot\alpha}{}^{\dot\gamma}\big[(\tilde\rho_j)^{CA}\big(abc\big)
+(\tilde\rho_j)^{AC}\big(acb\big)\big]
\col\\
V_{\bar cAc}
&=
\cvertr[
\fmfiv{label=$\scriptstyle a$,l.dist=2}{vloc(__v1)}
\fmfiv{label=$\scriptstyle \nu b$,l.dist=2}{vloc(__v2)}
\fmfiv{label=$\scriptstyle c$,l.dist=2}{vloc(__v3)}
]{dots_rar,label=$\scriptstyle p$,l.side=left,l.dist=2,l.dist=2}{photon,label=$\scriptstyle q$,l.side=right,l.dist=2,l.dist=2}{dots_ar,label=$\scriptstyle r$,l.side=right,l.dist=2,l.dist=2}
=-\frac{g_\YM}{\sqrt{2}}p^\nu\big(a\comm{b}{c}\big)
\col\\
\end{aligned}
\end{gathered}
\end{equation}
where we have used the abbreviation \eqref{trdef} for color traces.
Labels are read off clockwise 
starting with the leg in the upper left corner and all momenta are 
directed such that they leave the vertices. Thus, for particles the 
momenta are directed \emph{against} the R-charge flow that is indicated by the 
arrows on the lines.
The quartic vertices read
\begin{equation}\label{qvertices}
\begin{gathered}
\begin{aligned}
V_{AAAA}
=
\qvertr[
\fmfiv{label=$\scriptstyle \mu a$,l.dist=2}{vloc(__v1)}
\fmfiv{label=$\scriptstyle \nu b$,l.dist=2}{vloc(__v2)}
\fmfiv{label=$\scriptstyle \rho c$,l.dist=2}{vloc(__v3)}
\fmfiv{label=$\scriptstyle \sigma d$,l.dist=2}{vloc(__v4)}
]{photon,label=$\scriptstyle $,l.side=left,l.dist=2,l.dist=2}{photon,label=$\scriptstyle $,l.side=left,l.dist=2,l.dist=2}{photon,label=$\scriptstyle $,l.side=left,l.dist=2,l.dist=2}{photon,label=$\scriptstyle $,l.side=left,l.dist=2,l.dist=2}
&=
\smash[b]{{}\frac{g_\YM^2}{2}}
\big[(2g_{\mu\rho}g_{\nu\sigma}-g_{\mu\sigma}g_{\nu\rho}-g_{\mu\nu}g_{\rho\sigma})\big(\comm{a}{b}\comm{c}{d}\big)\\[-1.25\baselineskip]
&\hphantom{{}={}\frac{g_\YM^2}{\sqrt{2}}\big[}
+(2g_{\mu\nu}g_{\rho\sigma}-g_{\mu\sigma}g_{\nu\rho}-g_{\mu\rho}g_{\nu\sigma})\big(\comm{a}{c}\comm{b}{d}\big)
\big]
\col\\
V_{A\bar\phi A\phi}
=
\qvertr[
\fmfiv{label=$\scriptstyle \mu a$,l.dist=2}{vloc(__v1)}
\fmfiv{label=$\scriptstyle ib$,l.dist=2}{vloc(__v2)}
\fmfiv{label=$\scriptstyle \nu c$,l.dist=2}{vloc(__v3)}
\fmfiv{label=$\scriptstyle jd$,l.dist=2}{vloc(__v4)}
]{photon,label=$\scriptstyle $,l.side=left,l.dist=2,l.dist=2}{plain_rar,label=$\scriptstyle $,l.side=left,l.dist=2,l.dist=2}{photon,label=$\scriptstyle $,l.side=left,l.dist=2,l.dist=2}{plain_ar,label=$\scriptstyle $,l.side=left,l.dist=2,l.dist=2}
&=
\frac{g_\YM^2}{2}g_{\mu\nu}\delta^i_j\big[\big(\comm{a}{b}\comm{c}{d}\big)
+\big(\comm{a}{d}\comm{c}{b}\big)\big]
\col\\
V^{\text{D}}_{\bar\phi\phi\bar\phi\phi}
=
\qvertr[
\fmfiv{label=$\scriptstyle i a$,l.dist=2}{vloc(__v1)}
\fmfiv{label=$\scriptstyle j b$,l.dist=2}{vloc(__v2)}
\fmfiv{label=$\scriptstyle k c$,l.dist=2}{vloc(__v3)}
\fmfiv{label=$\scriptstyle l d$,l.dist=2}{vloc(__v4)}
]{plain_rar,label=$\scriptstyle $,l.side=left,l.dist=2,l.dist=2}{plain_ar,label=$\scriptstyle $,l.side=left,l.dist=2,l.dist=2}{plain_rar,label=$\scriptstyle $,l.side=left,l.dist=2,l.dist=2}{plain_ar,label=$\scriptstyle $,l.side=left,l.dist=2,l.dist=2}{}{}{}{}
&=
-\frac{g_\YM^2}{2}
\big[(\delta_j^i\delta_l^k+\delta_l^i\delta_j^k)\big(\big(abcd\big)+\big(adcb\big)\big)
\\[-1.25\baselineskip]
&\hphantom{{}={}-\frac{g_\YM^2}{2}\big[}
-\delta_j^i\delta_l^k\big(\big(abdc\big)+\big(acdb\big)\big)
-\delta_l^i\delta_j^k\big(\big(acbd\big)+\big(adbc\big)\big)\\
&\hphantom{{}={}-\frac{g_\YM^2}{2}\big[}
+\frac{4}{N}\big(Q^{ik}_{\text{D}\,jl}\big(ab\big)\big(cd\big)+Q^{ik}_{\text{D}\,lj}\big(ad\big)\big(cb\big)\big)
\big]
\col\\
V^{\text{F}}_{\bar\phi\bar\phi\phi\phi}
=
\qvertr[
\fmfiv{label=$\scriptstyle i a$,l.dist=2}{vloc(__v1)}
\fmfiv{label=$\scriptstyle j b$,l.dist=2}{vloc(__v2)}
\fmfiv{label=$\scriptstyle k c$,l.dist=2}{vloc(__v3)}
\fmfiv{label=$\scriptstyle l d$,l.dist=2}{vloc(__v4)}
]{plain_rar,label=$\scriptstyle $,l.side=left,l.dist=2,l.dist=2}{plain_rar,label=$\scriptstyle $,l.side=left,l.dist=2,l.dist=2}{plain_ar,label=$\scriptstyle $,l.side=left,l.dist=2,l.dist=2}{plain_ar,label=$\scriptstyle $,l.side=left,l.dist=2,l.dist=2}{}{}{}{}
&=
g_\YM^2
\Big[F^{ij}_{lk}\big(abcd\big)
+F^{ji}_{kl}\big(adcb\big)
+F^{ij}_{kl}\big(abdc\big)
+F^{ji}_{lk}\big(acdb\big)\\[-1.25\baselineskip]
&\phantom{{}={}g_\YM^2\Big(}
-\frac{s}{N}
(F^{ij}_{lk}+F^{ij}_{kl}+F^{jl}_{lk}+F^{ji}_{kl})\big(ab\big)\big(cd\big)\\
&\phantom{{}={}g_\YM^2\Big(}
-\frac{4}{N}Q^{ij}_{\text{F}\,kl}\big(ab\big)\big(cd\big)
\Big]
\col\\
\end{aligned}
\end{gathered}
\end{equation}
where we have split the quartic scalar interactions into those originating from D-terms and F-terms in the supersymmetric case. Moreover, we have kept the parameter $s$, which we set to its respective value $s=0$ and $s=1$ in the $U(N)$ and $SU(N)$ theory, cf.\ Section \ref{sec:doubletrcoupl}. We have also included the multi-trace couplings \eqref{dtc}, which are the only possible extension in the $SU(N)$ theory. The Feynman rules for the remaining multi-trace couplings of Section \ref{sec:multitrdef}, which can occur in the $U(N)$ theory, follow analogously.

The signs from permuting fermions within the Wick contractions are determined 
in analogy to the superspace case \cite{Gates:1983nr}:
\begin{enumerate}
\item Write down all factors from the vertices involving external (uncontracted)
spinor indices in the same ordering as they appear within the correlation function.
\item Write down all other factors involving spinor indices (e.g.\ propagators)
carefully keeping their internal ordering of indices, e.g.\  
$\alpha$ is left of $\dot\beta$ in $p_\alpha{}^{\dot\beta}$.
\item Eliminate $\delta_\alpha{}^\beta$, $\delta_{\dot\alpha}{}^{\dot\beta}$ and
bring contracted index pairs into canonical ordering, i.e.\ the index that 
is on the left side within the contracted pair is an upper index 
and the right one is a lower one.
\item Draw vertical parallel lines from the external indices downwards.
\item Connect contracted index pairs by lines. They cross the vertical 
lines and other lines of contracted index pairs.
Count the number $n$ of intersections of the lines and put a factor
$(-1)^n$ in front of the expression.
\item Reshuffle the product and change the up-down 
positions of contracted indices at your convenience, 
considering a factor $-1$ for each position-flip within contracted index pairs.
\end{enumerate}

\section{Tensor identities}
\label{app:tensorid}

In this appendix, we explicitly evaluate the tensor combinations that are encountered in the Feynman diagram analysis in Section \ref{sec:zombiekiller}. Recall that $i,j,k=1,2,3$ and $A,B,C=1,2,3,4$.

Introducing the transverse Kronecker delta
\begin{equation}
\tau_A^B=\delta_A^i\delta_i^B=\delta_A^B-\delta_A^4\delta_4^B\col
\end{equation}
we find the following auxiliary identities for certain contractions of the Yukawa-type couplings \eqref{rhodef}:
\begin{equation}
\begin{aligned}
(\rho_i\rho^{\dagger\,j})_A{}^B
&=(\rho_{i})_{AC}(\rho^{\dagger\,j})^{CB}
=\sum_{C\neq A,B,i,j,4}(\delta_i^j\tau_A^B-\delta_A^j\delta_i^B\e^{\frac{i}{2}(\Gamma_{AC}-\Gamma_{BC})})\col\\
((\rho^{\dagger\,i})^{\T}\rho_j)
&=(\rho_i(\rho^{\dagger\,j})^{\T})^\ast
\col\\
(\rho_i(\rho^{\dagger\,j})^{\T})_A{}^B
&=(\rho_{i})_{AC}(\rho^{\dagger\,j})^{BC}
=-\sum_{C\neq A,B,i,j,4}(\delta_i^j\tau_A^B\e^{i\Gamma_{AC}}-\delta_A^j\delta_i^B\e^{\frac{i}{2}(\Gamma_{AC}+\Gamma_{BC})})
\col\\
(\tilde\rho_i\tilde\rho^{\dagger\,j})^A{}_B
&=(\tilde\rho_i)^{AC}(\tilde\rho^{\dagger\,j})_{CB}
=\delta_4^A\delta_B^4\delta_i^j+\delta^A_i\delta^j_B\e^{\frac{i}{2}(\Gamma_{i4}-\Gamma_{j4})}\col\\
((\tilde\rho^{\dagger\,i})^{\T}\tilde\rho_j)
&=(\tilde\rho_i(\tilde\rho^{\dagger\,j})^{\T})^\ast
\col\\
(\tilde\rho_i(\tilde\rho^{\dagger\,j})^{\T})^A{}_B
&=(\tilde\rho_i)^{AC}(\tilde\rho^{\dagger\,j})_{BC}
=-\delta_4^A\delta_B^4\delta_i^j\e^{i\Gamma_{4i}}
-\delta_i^A\delta_B^j\e^{\frac{i}{2}(\Gamma_{i4}+\Gamma_{j4})}
\pnt
\end{aligned}
\end{equation}
With these results, the traces of two Yukawa coupling tensors that appear in the 
one-loop self-energies evaluate to
\begin{equation}
\begin{aligned}\label{rhorhotr}
\tr(\rho_i\rho^{\dagger\,j})
&=(\rho_{i})_{AC}(\rho^{\dagger\,j})^{CA}
=\sum_{A\neq i,4}\sum_{C\neq A,i,4}(\delta_i^j\tau_A^B-\delta_A^j\delta_i^A)
=2\delta_i^j\col\\
\tr(\rho_i(\rho^{\dagger\,j})^{\T})
&=(\rho_{i})_{AC}(\rho^{\dagger\,j})^{AC}
=-\sum_{A\neq i,4}\sum_{C\neq A,i,4}\delta_i^j\e^{i\Gamma_{AC}}
=-2\delta_i^j\cos\tfrac{1}{2}\epsilon_{ikl}(\gamma_k+\gamma_l)
\col\\
\tr(\tilde\rho_i\tilde\rho^{\dagger\,j})
&=(\tilde\rho_i)^{AC}(\tilde\rho^{\dagger\,j})_{CA}
=\sum_{A}(\delta_4^A\delta_A^4\delta_i^j+\delta^A_i\delta^j_A)=2\delta_i^j\col\\
\tr(\tilde\rho_i(\tilde\rho^{\dagger\,j})^{\T})
&=(\tilde\rho_i)^{AC}(\tilde\rho^{\dagger\,j})_{AC}
=-\delta_i^j\e^{i\Gamma_{4i}}
-\delta_i^j\e^{i\Gamma_{i4}}
=-2\delta_i^j\cos\tfrac{1}{2}\epsilon_{ikl}(\gamma_k-\gamma_l)
\pnt
\end{aligned}
\end{equation}
Once again, Einstein's summation convention should not be applied on the rightmost sides of the equations. Instead, the index $i=1,2,3$ is fixed and $j$ and $k$ assume the values of the corresponding cyclic permutation $(i,j,k)\in\{(1,2,3),(2,3,1),(3,1,2)\}$.
For the evaluation of the fermion-box contribution to the renormalization of the double-trace couplings \eqref{zombiekiller}, we need some traces of four Yukawa coupling tensors with identical (not summed) bosonic index $i$:
\begin{equation}
\begin{aligned}\label{l4rhotraces}
\tr\big[\rho_i(\rho^{\dagger i})^{\T}(\tilde\rho^{\dagger i})^{\T}\tilde\rho_i\big]
&=0
\col\\
\tr\big[\tilde\rho_i(\tilde\rho^{\dagger i})^{\T}(\rho^{\dagger i})^{\T}\rho_i\big]
&=0
\col\\
\tr\big[\rho_i(\rho^{\dagger i})^{\T}\rho_i(\rho^{\dagger i})^{\T}\big]
&=\sum_{A,C\neq i,4}\e^{2i\Gamma_{AC}}
=\sum_{A,C\neq i,4}\cos 2\Gamma_{AC}
=2\cos\epsilon_{ijk}(\gamma_j+\gamma_k)
\col\\
\tr\big[\tilde\rho_i(\tilde\rho^{\dagger i})^{\T}\tilde\rho_i(\tilde\rho^{\dagger i})^{\T}\big]
&=\e^{2i\Gamma_{4i}}+\e^{2i\Gamma_{i4}}=2\cos 2\Gamma_{4i}
=2\cos\epsilon_{ijk}(\gamma_j-\gamma_k)
\pnt
\end{aligned}
\end{equation}

The one-loop interaction of four scalars with identical field flavors via two F-term-type interactions requires the evaluation of the following expression
\begin{equation}
\begin{aligned}
\sum_{r=1}^3F^{ir}_{ri}F^{ir}_{ri}
&=2+\sum_{r=1}^3(2+Q^{ir}_{ri})Q^{ir}_{ri}
=2\sum_{\substack{r=1 \\r\neq i}}^3\e^{i\Gamma^+_{ir}}\cos\Gamma^+_{ir}-2
\col\\
\end{aligned}
\end{equation}
which, using \eqref{Fbarrel}, immediately yields
\begin{equation}
\begin{aligned}
\sum_{r=1}^3F^{ri}_{ir}F^{ri}_{ir}
&=\sum_{r=1}^3(F^{ir}_{ri}F^{ir}_{ri})^\ast
=2\sum_{\substack{r=1 \\r\neq i}}^3\e^{-i\Gamma^+_{ir}}\cos\Gamma^+_{ir}-2\pnt
\end{aligned}
\end{equation}
For the combined sums of the first two lines in \eqref{diagres}, we hence obtain the result
\begin{equation}
\begin{aligned}\label{FFsum}
\sum_{r=1}^3(F^{ir}_{ri}F^{ir}_{ri}+F^{ri}_{ir}F^{ri}_{ir})
&=4\sum_{\substack{r=1 \\r\neq i}}^3\cos^2\Gamma^+_{ir}-4
=4(\cos^2\epsilon_{ijk}\gamma_j+\cos^2\epsilon_{ijk}\gamma_k-1)
\pnt
\end{aligned}
\end{equation}

\section{One-loop self-energies}
\label{app:oneloopse}

Using the relations \eqref{rhorhotr}, the UV divergences of the one-loop 
self-energy contributions to the scalar propagators 
are determined as
\begin{equation}
\begin{aligned}\label{sephi}
\Kop\Big[\swfone[
\fmfiv{label=$\scriptstyle i a$,l.dist=2}{vloc(__v1)}
\fmfiv{label=$\scriptstyle j b$,l.dist=2}{vloc(__v2)}
]{plain_ar}{dashes_ar,left=1}{dashes_rar,left=1}
\Big]
&=-2p^2\frac{g_\YM^2}{(4\pi)^2\varepsilon}\delta_i^j\big[N\big(ab\big)-\cos\tfrac{1}{2}\epsilon_{ikl}(\gamma_k-\gamma_l)\big(a\big)\big(b\big)\big]
\col\\
\Kop\Big[\swfone[
\fmfiv{label=$\scriptstyle i a$,l.dist=2}{vloc(__v1)}
\fmfiv{label=$\scriptstyle j b$,l.dist=2}{vloc(__v2)}
]{plain_ar}{dashes_rar,left=1}{dashes_ar,left=1}
\Big]
&=-2p^2\frac{g_\YM^2}{(4\pi)^2\varepsilon}\delta_i^j\big[N\big(ab\big)
-\cos\tfrac{1}{2}\epsilon_{ikl}(\gamma_k+\gamma_l)\big(a\big)\big(b\big)\big]
\col\\
\Kop\Big[\swfone[
\fmfiv{label=$\scriptstyle i a$,l.dist=2}{vloc(__v1)}
\fmfiv{label=$\scriptstyle j b$,l.dist=2}{vloc(__v2)}
]{plain_ar}{photon,left=1}{plain_rar}
\Big]
&=p^2\frac{g_\YM^2}{(4\pi)^2\varepsilon}\delta_i^j(3-\alpha)\big[N\big(ab\big)-\big(a\big)\big(b\big)\big]
\col
\end{aligned}
\end{equation}
for external momentum $p^\nu$. The single-trace coefficient of the sum of the expressions yields the counter term of the wave function renormalization for the $SU(N)$ fields. It reads 
\begin{equation}
\begin{aligned}
\label{deltaphi}
\delta_{\phi^i}&=\frac{1}{p^2}\Kop\Big[
\settoheight{\eqoff}{$\times$}%
\setlength{\eqoff}{0.5\eqoff}%
\addtolength{\eqoff}{-7.5\unitlength}%
\raisebox{\eqoff}{%
\fmfframe(3,0)(3,0){%
\begin{fmfchar*}(20,15)
\fmfleft{v1}
\fmfright{v2}
\fmf{plain_ar}{v1,vl}
\fmf{phantom}{vl,vr}
\fmf{plain_ar}{vr,v2}
\fmffreeze
\fmfposition
\fmfiv{label=$\scriptstyle i a$,l.dist=2}{vloc(__v1)}
\fmfiv{label=$\scriptstyle j b$,l.dist=2}{vloc(__v2)}
\fmfcmd{pair vert[]; vert1 = vloc(__vl);}
\vacpolp[1]{vert1--vloc(__vr)}
\end{fmfchar*}}}
\Big]\Big|_{\delta_i^j(ab)}
&=-\frac{g_\YM^2N}{(4\pi)^2\varepsilon}(1+\alpha)
\pnt
\end{aligned}
\end{equation}

\section{\texorpdfstring{Coupling renormalization and $\beta$-functions}{Coupling renormalization and beta-functions}}
\label{app:couplren}

In this appendix, we review the definition of the $\beta$-functions and how they are obtained from renormalized couplings and fields. The coupling \eqref{zombiekiller} written in terms of bare coupling constants and bare fields has to be identified with \eqref{zombiekillerct}, i.e.\ the respective coupling and counter term in renormalized perturbation theory in $D=4-2\varepsilon$ dimensions. This yields
\begin{equation}\label{zombiekillerrel}
-\frac{g_{\YM\,0}^2}{N}Q_{0\,\text{F}\,ii}^{ii}\tr(\bar\phi_{0\,i}\bar\phi_{0\,i})\tr(\phi_0^i\phi_0^i)
=-\frac{\mu^{2\varepsilon}g_\YM^2}{N}(Q_{\text{F}\,ii}^{ii}+\delta Q_{\text{F}\,ii}^{ii})\tr(\bar\phi_i\bar\phi_i)\tr(\phi^i\phi^i)\col
\end{equation}
where we have introduced the 't Hooft mass $\mu$ that rescales 
the unrenormalized Yang-Mills coupling $g_{\YM\,0}=\mu^\varepsilon g_\YM$.
The renormalized couplings and fields are given in terms of the renormalization 
constants and bare quantities as\footnote{As in \eqref{zombiekillerrel}, the
scalar field is restricted to its $SU(N)$ components.}
\begin{equation}
\begin{aligned}\label{Qphiren}
Q_{\text{F}\,ii}^{ii}&=\mathcal{Z}_{Q_{\text{F}\,ii}^{ii}}^{-1}Q_{0\,\text{F}\,ii}^{ii}
\col\qquad
&\mathcal{Z}_{Q_{\text{F}\,ii}^{ii}}&=1+\delta_{Q_{\text{F}\,ii}^{ii}}\col\\
\phi^i&=\mathcal{Z}_{\phi^i}^{-\frac{1}{2}}\phi_0^i
\col\qquad
&\mathcal{Z}_{\phi^i}&=1+\delta_{\phi^i}\pnt
\end{aligned}
\end{equation}
Inserting these expressions into \eqref{zombiekillerrel}, we obtain
\begin{equation}
\mathcal{Z}_{Q_{\text{F}\,ii}^{ii}}
=(1+(Q_{\text{F}\,ii}^{ii})^{-1}\delta Q_{\text{F}\,ii}^{ii}\big)\mathcal{Z}_{\phi^i}^{-2}
\pnt
\end{equation}
At leading order in the coupling constant, this yields
\begin{equation}\label{deltaQdef}
\delta_{Q_{\text{F}\,ii}^{ii}}=\frac{1}{Q_{\text{F}\,ii}^{ii}}(\delta Q_{\text{F}\,ii}^{ii}-2Q_{\text{F}\,ii}^{ii}\delta_{\phi^i})
\pnt
\end{equation}
The $\beta$-functions are defined as
\begin{equation}
\beta_{g_\YM}=\mu\deriv{\mu}g_{\YM}\col\qquad
\beta_{Q_{\text{F}\,ii}^{ii}}=\mu\deriv{\mu}Q_{\text{F}\,ii}^{ii}
\pnt
\end{equation}
The independence of the bare coupling constants from $\mu$ implies 
the following relations
\begin{equation}
\begin{aligned}
0&=\mu\deriv{\mu}g_{\YM\,0}=\Big(\mu\parderiv{\mu}+\beta_{g_\YM}\parderiv{g_\YM}\Big)\mu^\varepsilon g_\YM=
\mu^\varepsilon(\epsilon g_\YM+\beta_{g_\YM})\col\\
0&=\mu\deriv{\mu}Q_{0\,\text{F}\,ii}^{ii}
=Q_{\text{F}\,ii}^{ii}\Big(\beta_{g_\YM}\parderiv{g_\YM}+\beta_{Q_{\text{F}\,ii}^{ii}}\parderiv{Q_{\text{F}\,ii}^{ii}}\Big)\mathcal{Z}_{Q_{\text{F}\,ii}^{ii}}+\mathcal{Z}_{Q_{\text{F}\,ii}^{ii}}\beta_{Q_{\text{F}\,ii}^{ii}}
\pnt
\end{aligned}
\end{equation}
The first equation determines the $\beta$-function for 
$g_\YM$, 
\begin{equation}
\beta_{g_\YM}=-\varepsilon g_{\YM}
\col
\end{equation}
which vanishes in the four-dimensional theory, i.e.\ for $\varepsilon=0$. This is expected since $g_\YM$ is not renormalized. Inserting this result, the second equation determines the $\beta$-function for the coupling $Q_{\text{F}\,ii}^{ii}$ as
\begin{equation}
\begin{aligned}
0
&=Q_{\text{F}\,ii}^{ii}\Big(-\varepsilon g_\YM\parderiv{g_\YM}+\beta_{Q_{\text{F}\,ii}^{ii}}\parderiv{Q_{\text{F}\,ii}^{ii}}\Big)\ln\mathcal{Z}_{Q_{\text{F}\,ii}^{ii}}+\beta_{Q_{\text{F}\,ii}^{ii}}
\pnt
\end{aligned}
\end{equation}
At lowest order, where the second term in parentheses does not contribute, we find after inserting \eqref{Qphiren} and \eqref{deltaQdef}
\begin{equation}
\begin{aligned}\label{betaQlo}
\beta_{Q_{\text{F}\,ii}^{ii}}
&=Q_{\text{F}\,ii}^{ii}\varepsilon g_\YM\parderiv{g_\YM}\ln\mathcal{Z}_{Q_{\text{F}\,ii}^{ii}}
=\varepsilon g_\YM\parderiv{g_\YM}Q_{\text{F}\,ii}^{ii}\delta_{Q_{\text{F}\,ii}^{ii}}
\pnt
\end{aligned}
\end{equation}

\footnotesize
\bibliographystyle{JHEP}
\bibliography{references}

\end{fmffile}
\end{document}